\documentclass[10pt,journal,compsoc]{./IEEEtran}

\usepackage{amsmath,amsfonts}
\usepackage{algorithm}
\usepackage{array}
\usepackage[caption=false,font=normalsize,labelfont=sf,textfont=sf]{subfig}
\usepackage{textcomp}
\usepackage{stfloats}
\usepackage{url}
\usepackage{verbatim}
\usepackage{graphicx}
\usepackage{cite}
\usepackage{algpseudocode}
\usepackage{multirow}

\usepackage{pifont}
\usepackage{threeparttable}
\usepackage{booktabs}
\usepackage{ragged2e}
\usepackage{makecell}
\usepackage[justification=centering]{caption}

\begin{document}

\title{Facial Data Minimization: Shallow Model as Your Privacy Filter}

\author{Yuwen Pu\IEEEauthorrefmark{1}, Jiahao Chen\IEEEauthorrefmark{1}, Jiayu Pan, Diqun Yan, Xuhong Zhang, Shouling Ji\IEEEauthorrefmark{2}
\IEEEcompsocitemizethanks{
\IEEEcompsocthanksitem Y. Pu, J.Chen, S.Ji are with the College of Computer Science and Technology at Zhejiang University, Hangzhou, Zhejiang, 310027, China. E-mail:\{yw.pu, , sji\}@zju.edu.cn\protect\\
\IEEEcompsocthanksitem J. Pan, X.Zhang are with the College of Software Technology at Zhejiang University, NingBo, Zhejiang, 315048, China. E-mail: 13957897451@163.com, zhangxuhong@zju.edu.cn\protect\\
\IEEEcompsocthanksitem D. Yan is with the Faculty of Electrical Engineering and Computer Science at Ningbo University, NingBo, Zhejiang, 315211, China. E-mail: yandiqun@nbu.edu.cn\protect\\
\IEEEcompsocthanksitem \IEEEauthorrefmark{1} Y.Pu and J.Chen are the co-first authors.\protect\\
\IEEEcompsocthanksitem \IEEEauthorrefmark{2} S.Ji is the corresponding author.\protect\\
}
}





\IEEEtitleabstractindextext{
\begin{abstract}
Face recognition service has been used in many fields and brings much convenience to people. However, once a user's facial data is transmitted to a service provider, the user will lose control of his/her private data. In recent years, there are various security and privacy issues due to the leakage of facial data. Although many privacy-preserving methods have been proposed, they usually fail when they are not accessible to adversaries' strategies or the complete face recognition model. Hence, in this paper, by fully considering two cases of uploading facial images and facial features, which are very typical in practical face recognition service systems, we proposed a data privacy minimization transformation (PMT) method. This method can process the original facial data based on the shallow model of the authorized face recognition model to obtain the obfuscated data. The obfuscated data cannot only maintain satisfactory performance on authorized models and restrict the performance on other unauthorized models but also prevent original privacy data from leaking by AI methods and human visual theft. Additionally, since a service provider may execute preprocessing operations on the received data, we also propose an enhanced perturbation method to improve the robustness of PMT. Besides, to authorize one facial image to multiple service models simultaneously, a multiple-restriction mechanism is proposed to improve the scalability of PMT. Finally, we conduct extensive experiments and evaluate the effectiveness of the proposed PMT in defending against face reconstruction, data abuse, and face attribute estimation attacks. These experimental results demonstrate that PMT performs well in preventing facial data abuse and privacy leakage while maintaining high face recognition accuracy.
\end{abstract}
}

\maketitle

\section{Introduction}

\IEEEPARstart{F}ace recognition service has been used in many fields, such as security, finance, smart health, etc. It brings great convenience to individuals and enterprises. However, there are also many privacy and copyright issues due to facial data leakage. Over the past years, there are numerous incidents of data leakage exposed in various companies\cite{Drapkin23,Sullivan21}, sparking concerns among individuals about their privacy. For example, the New York Times reported that \textit{Clearview.ai} collected more than 3 billion online facial images without authentication\cite{Hill21}. These instances indicate that private facial data leakage is almost unavoidable. Once a user's facial data is directly uploaded to service provider, the attackers may conduct face attribute estimation\cite{zhong2016face,ding2018deep}, malicious editing\cite{Nehamas21,Gözükara23}, and other privacy-infringing operations without the user's authorization, and even stealthily sell the user's privacy data to others for benefits (e.g., some companies may require many facial data to train a face attribute analysis model for advertising recommendation). Thees data leakage and abuse issues described above severely threaten users' facial data privacy. 

For protecting users' facial data in face recognition services, some researchers have proposed to generate fake facial images with visually different identities by using generative adversarial networks\cite{qiu2021synface,yuan2022pro}, which achieved high recognition accuracy and defense performance. However, these methods can only work under a white-box setting. Some other researchers adopt adversarial strategy to defend against unauthorized use of private data\cite{zhong2022opom, yang2021towards, chandrasekaran2021face, hu2022protecting, shan2020fawkes, cherepanova2021lowkey}, but these methods lead to the performance degradation of the authorized service and suffer from low transferability and robustness\cite{deb2023faceguard}. 

\begin{figure}
    \centering
    \includegraphics[width=8cm]{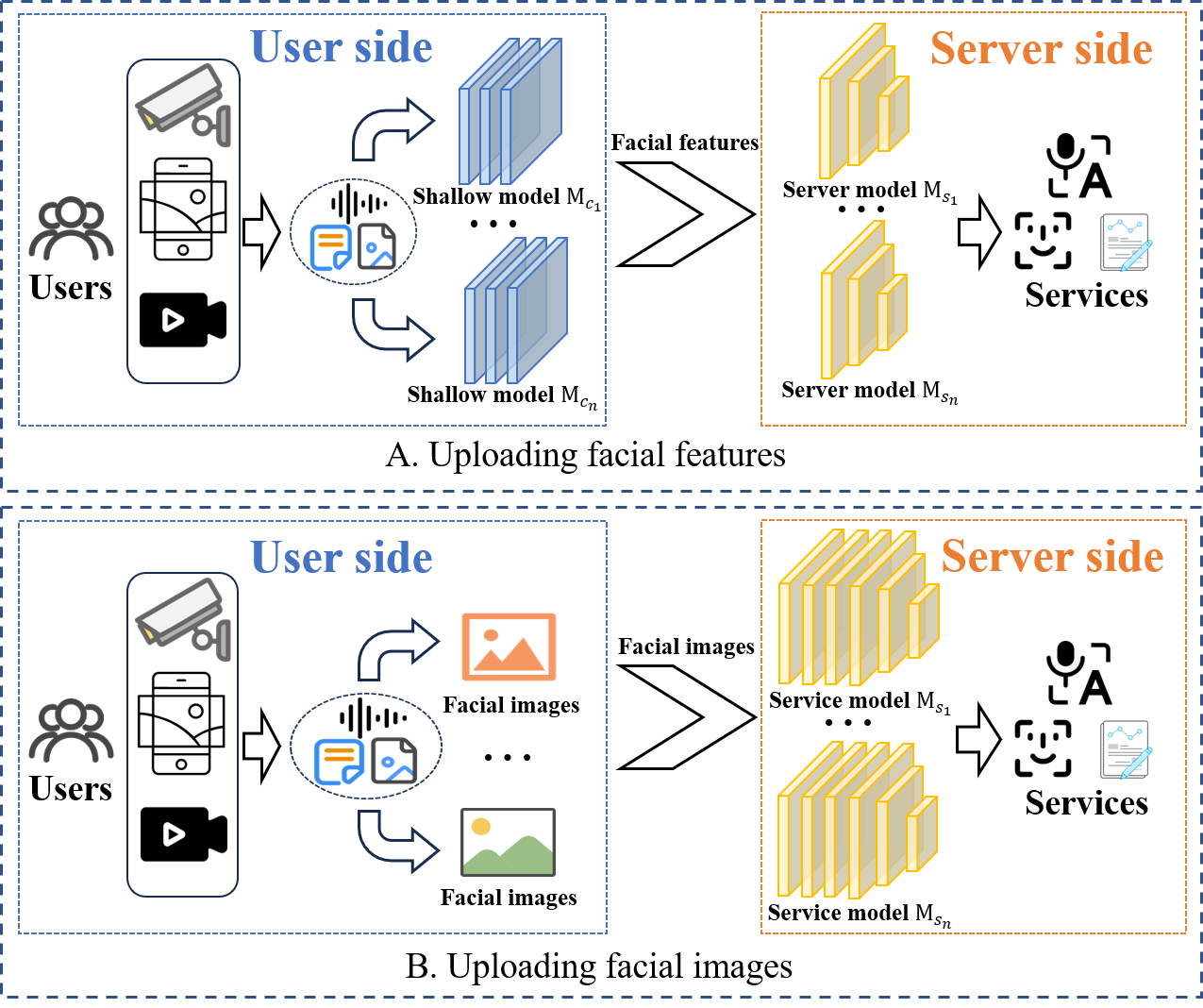}
    \caption{Overview of the Face Recognition System Architecture.}
    \label{fig:multics}
\end{figure}

To avoid privacy leakage caused by the original facial image transmission, several researchers proposed to upload facial features rather than facial images to the service provider\cite{wang2023privacy,li2021deepobfuscator}. That is, the service provider splits the face recognition network into a shallow model (the front layers of the network) and a server model (the remaining layers). Then, it shares and deploys the shallow model on the user side. When a user enjoys face recognition service, the user can extract shallow features from facial images based on the shallow model. Subsequently, these shallow features are transmitted to the service provider for further feature extraction. The above process is shown in Fig.\ref{fig:multics}. A. However, even when the service provider only has access to the user's facial features, he/she can still conduct a face reconstruction attack to recover the user's original facial images \cite{erdougan2022unsplit, mai2018reconstruction, fredrikson2015model, razzhigaev2020black, he2019model}. Firstly, the service provider is aware of the shallow model on the user side, enabling him/her to perform data inversion in a white-box scenario using the same shallow model\cite{erdougan2022unsplit, fredrikson2015model, razzhigaev2020black}. Additionally, the attacker can train a generative model to learn the mapping relationships from features to images by utilizing facial image and feature pairs\cite{zhmoginov2016inverting, mai2018reconstruction, he2019model}. In this way, during shallow model deployment, the user's data can be reconstructed based on the user's facial features. The above two types of data reconstruction attacks pose a severe threat to user privacy.

In order to defend against the above reconstruction attacks, various methods are proposed. \cite{abdalla2015simple, kou2021efficient, gentry2011implementing} suggest performing model training and inference in an encrypted space to ensure content obfuscation. However, this approach brings significant computational overhead and a certain degree of accuracy reduction. Adversarial perturbation-based methods\cite{wang2023privacy,li2021deepobfuscator} aim to add small adversarial perturbations to facial features or images, thereby causing confusion in malicious face editing or attribute inference applications. However, the generalization performance of these methods immensely depends on the transferability of adversarial samples since they require a surrogate model as a part of the perturbation optimization process. Therefore, when facing numerous unknown attacks, adding tiny perturbations to achieve privacy protection becomes impractical. Some methods\cite{yangprivatefl, chamikara2020privacy, Mao2018APD} attempt to protect facial data by using differential privacy techniques to add perturbations to the facial features. Nevertheless, it is challenging to balance between privacy-preserving and task accuracy. There are also some methods\cite{mi2022duetface, Wang2022PrivacyPreservingFR} that utilize frequency domain features for face recognition while achieving privacy-preserving goals. However, recent research points out that these approaches cannot defend against powerful reconstruction attacks \cite{wang2023privacy}. 

Besides, existing methods mainly aim to process a facial image or feature for only one specific service model so as to achieve privacy-preserving or data abuse prevention. However, in a real-world scenario, one user may simultaneously request multiple face recognition services from different service providers. These circumstances require that users generate different facial images or features for all authorized service models and upload them to corresponding service providers. It brings much computation cost for the user side. 

In summary, there are various privacy-preserving methods for face recognition services, but they are not practical. In particular, existing methods only consider one scenario (uploading facial images or facial features) and cannot achieve both privacy-preserving and data abuse prevention. Moreover, they cannot simultaneously authorize one facial image to multiple service models to prevent data abuse. Besides, they usually fail to protect user privacy when they are not accessible to adversaries’ strategies or the complete face recognition model, which is impractical because adversaries’ strategies are diverse and the face recognition model is private for service providers. Hence, in this paper, we aim to overcome the weaknesses of the existing methods and propose a plug-and-play method. Undoubtedly, it is non-trivial to realize this objective. There are three main challenges that need to be addressed as follows. 

\begin{itemize}
\item Once users' facial images or features are transmitted to the service provider, how can users gain more control over the usage of their facial data by authorizing operations?
\item How to achieve facial data privacy-preserving while maintaining the accuracy of the main task? That is, how to find a delicate balance between privacy-preserving and data utility when handling the original facial data?
\item How to ensure the generalization capability of privacy-preserving methods in the face of various unknown attacks?
\end{itemize}

To address the above challenges, we propose a privacy minimization transformation (PMT) method to implement data authorization and privacy-preserving. The intuition of PMT is to minimize the distance between features extracted from the original facial data and the features extracted from the obfuscated data by the shallow model, thereby obtaining the final obfuscated data after optimization. We further proposed an enhanced perturbation approach and a multiple-restriction mechanism to improve the robustness and scalability of PMT. In our method, the service provider needs to share the shallow model (the front layers of the entire service model) with the registered users like a public key. Then, the users can compute the obfuscated data based on PMT and the shallow model. Subsequently, the obfuscated facial images or features are handed over to the service provider for further face recognition service. Even if an attacker obtains the obfuscated facial images and features, he/she cannot disclose the privacy of the original data by conducting a face reconstruction attack or attribute estimation attack. Moreover, these obfuscated facial images can only maintain satisfactory performance on the authorized models and restrict the performance on other unauthorized models. In particular, one obfuscated facial image can be authorized to multiple facial service models to prevent data abuse. 

Our main contributions are summarized as follows:
\begin{itemize}
\item
We propose a novel privacy minimization transformation method, which can achieve privacy-preserving and data authorization by minimizing the distance between the features extracted from the original facial data and those extracted from the obfuscated data. Besides, PMT does not require complete face recognition model or the knowledge of the attack strategy except for the shallow model of face recognition service.
\item
Considering the various preprocessing operations that may be executed on the received data by service providers in real-world scenarios, an enhanced perturbation mechanism is proposed to improve the robustness of PMT.
\item
Given that one facial image may be delegated to multiple service models simultaneously, a multiple-restriction mechanism is proposed to improve the scalability of PMT.
\item
We conduct comprehensive empirical studies on different datasets and models. The experimental results show that our method can achieve a great trade-off between privacy-preserving and data utility regardless of the service providers' model architecture. Moreover, it is also effective in restricting the performance of other unauthorized models while containing no degradation performance in the authorized model.
\end{itemize}

\section{Related Works}

In this section, we review the relevant works of face reconstruction and attribute estimation. Then, the development of privacy-preserving methods for face recognition is introduced.

\subsection{Face Reconstruction}
Traditional face reconstruction uses regression or affine transformation-based methods to reconstruct the face in the early stage \cite{mignon2013reconstructing,mohanty2007scores}. However, they are not applicable when features are extracted by deep learning models. Consequently, by solving an optimization problem, Fredrikson et al.\cite{fredrikson2015model} and Razzhigaev et al.\cite{razzhigaev2020black} reconstructed facial images based on the features extracted by deep learning models. More specifically, they minimized the distance between the features of reconstructed and target images. The basic workflow of the face reconstruction attack is shown in Fig.\ref{fig:attack}. Note that many methods have been proposed to boost the performance of data reconstruction.
There are generally two categories for face reconstruction, including optimization-based and model-based methods. The former requires the adversary's accessibility to a surrogate model that has a similar functionality to the victim model. Erdougan et al.\cite{erdougan2022unsplit} recovered the input samples and optimized a functionally similar model to the client model simultaneously. In recent years, the development of model inversion attacks \cite{zhang2020secret, chen2021knowledge, struppek2022plug, papernot2017practical, juuti2019prada, orekondy2019knockoff} has boosted the performance of data reconstruction by optimization-based methods. However, these optimization-based methods require a significant amount of time or involve querying tens of thousands of times. Therefore, a model-based method was proposed to reconstruct the facial images given the facial features\cite{cole2017synthesizing,dosovitskiy2016inverting,he2019model,mai2018reconstruction,zhmoginov2016inverting}. The latter utilizes facial images and feature pairs to train a generative model to learn the mapping relationships between features and images. \cite{dosovitskiy2016inverting, he2019model, mai2018reconstruction, zhmoginov2016inverting} involve training a new face reconstruction model using datasets that have a similar distribution to the test set of face recognition models. However, the performance of this kind of method is limited to the distribution differences between the collected pairs and those in practice. In this paper, we adopt both types of reconstruction attacks to validate the effectiveness of the proposed method.

\subsection{Face Attribute Estimation}
\begin{figure}
    \centering
    \includegraphics[width=8.7cm]{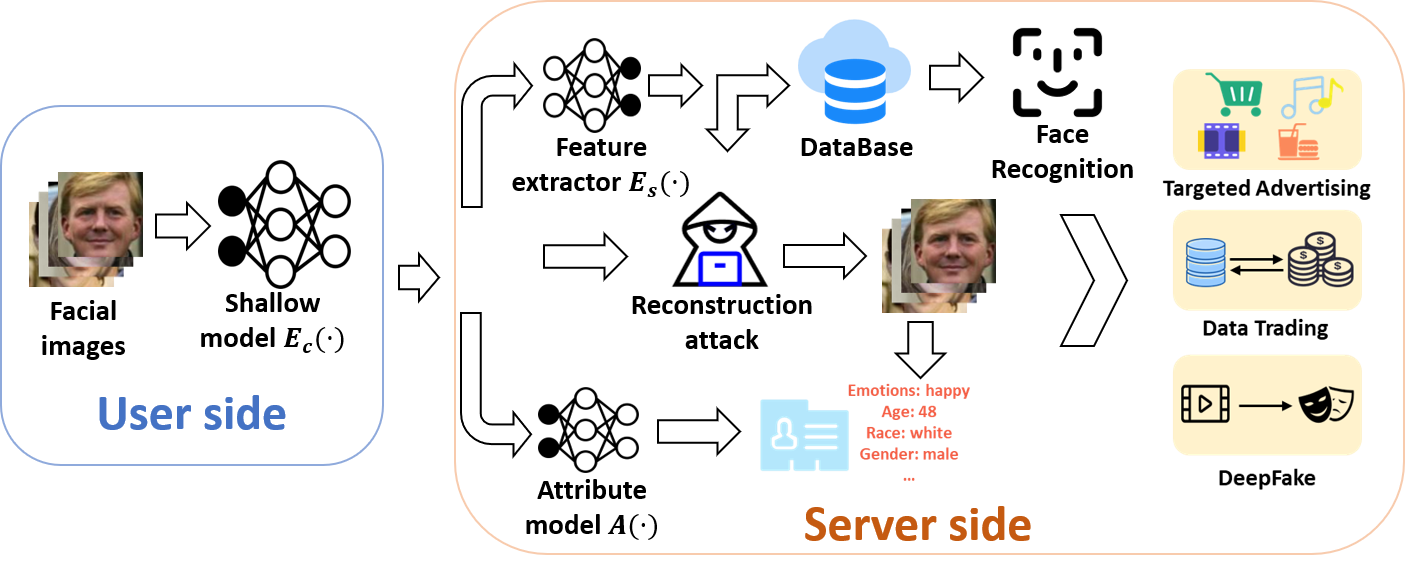}
    \caption{Overview of various attacks.}
    \label{fig:attack}
\end{figure}

Human face conveys vital signals for social interaction, encompassing a diverse range of significant details, such as the individual's demographic attributes (age, gender, and race), hairstyle, clothing, and more\cite{han2017heterogeneous}. Recently, many relevant works \cite{liu2015deep,zhong2016face,ding2018deep} have been proposed. Given facial images, face attribute estimation aims to estimate the attributes of the face owners using a model, as shown in Fig. \ref{fig:attack}. As mentioned before, real-world face recognition services (e.g., Azure, Megvii, Aliyun) often need to upload the original facial images. Hence, the server may potentially utilize these techniques without authorization to infer and analyze the attributes of users' facial images, thereby seriously infringing upon their privacy. Even without the original facial images, the attackers can follow the aforementioned reconstruction techniques to obtain the facial images and conduct face attribute estimation. In this paper, we use the open-source toolbox DeepFace\cite{deepface} to complete face attribute estimation (including age, gender, facial expression, and race) as an evaluation for the proposed PMT.
\begin{table*}[]
\begin{center}
\renewcommand{\arraystretch}{1.5}
\caption{A comparison of facial privacy protection methods.}
\label{tab:comparision_work}
\begin{threeparttable}
\begin{tabular}{cccccccc}
\midrule
\textbf{Work}& \textbf{Modify}& \textbf{Knowledge}& \textbf{Strategy}& \textbf{Extra Model}& \textbf{Generic Defense}& \textbf{Utility}& \textbf{Privacy} \\ 
\hline
Hu et al.\cite{hu2022protecting} &\ding{56} &Model,Data &Adversarial inputs &\ding{52} &\ding{56} &$\star$ &$\star$\\

Wang et al.\cite{wang2023privacy} &\ding{56} &Model,Data &Adversarial features &\ding{52} &\ding{56} &$\star\star$ &$\star\star$\\

Zhong et al.\cite{zhong2022opom} &\ding{56} &Model &Adversarial inputs &\ding{56} &\ding{56} &$\star$ &$\star$\\

Chamikara et al.\cite{chamikara2020privacy} &\ding{52}&Training control&Differential privacy&\ding{56}&\ding{52}&$\star$&$\star\star$\\

Li et al.\cite{li2021deepobfuscator}& \ding{52}& Training control& Adversarial learning& \ding{52}& \ding{56}& $\star\star$& $\star$\\

Mi et al.\cite{mi2022duetface}& \ding{56}& Training control&Frequency domain&\ding{56}& \ding{52}& $\star$& $\star\star$\\

Wang et al.\cite{wang2022privacy}& \ding{52}&Training control&Frequency domain&\ding{56}&\ding{52}&$\star\star$&$\star\star$\\

\hline
Our PMT & \ding{56}& Shallow model& Feature approximation& \ding{56}& \ding{52}& $\star\star\star$& $\star\star\star$\\
\midrule
\end{tabular}

 \begin{tablenotes}   
    \footnotesize               
    \item[1] \textbf{Modify} means whether the service model needs to be modified compared with a normally trained model. By obviating this limitation, our PMT is generic to serve as a plug-and-play defense.
    \item[2] \textbf{Knowledge} is the defender's knowledge that is necessary for making defense procedures.
    \item[3] \textbf{Strategy} is the specific and key method used by the defender to prevent privacy leakage.
    \item[4] \textbf{Extra Model} means whether the defender needs to train an extra model.
    \item[5] \textbf{Generic Defense} means whether this method is a generic defense to various attacks (including reconstruction attack, data abuse and attribute estimation attack).
    \item[6] \textbf{Utility} and \textbf{Privacy} stand for the face recognition accuracy and the performance of the privacy-preserving method, respectively.
\end{tablenotes}
\end{threeparttable}
\end{center}
\end{table*}

\subsection{Face Privacy Protection}
As privacy concerns have gained increasing attention, researchers have presented a wide range of privacy protection methods in recent years. This paper provides a broad classification of these methods, dividing them into four main categories. Adversarial example-based methods generate adversarial perturbations or make style transfer for anonymizing the corresponding facial images \cite{shan2020fawkes, cherepanova2021lowkey,hu2022protecting,zhong2022opom} while maintaining image quality. However, this kind of method fails when the target models are able to make robust predictions \cite{radiya2021data} if users’ perturbed pictures are published and scraped for model training. Similarly, differential privacy-based approaches add crafted noises to the images or output to achieve privacy protection \cite{yangprivatefl}. However, the model performance is significantly affected \cite{chamikara2020privacy,Mao2018APD}. Encryption-based methods use techniques like functional encryption \cite{abdalla2015simple}, matrix encryption \cite{kou2021efficient}, and homomorphic encryption \cite{gentry2011implementing} to map the facial images and features into an encrypted space to safeguard the privacy of the data. However, these methods suffer from high computational costs. Model-based methods utilize unique model structures or training skills to avoid direct personal information leakage. For example, DuetFace \cite{mi2022duetface} and PPFR-FD\cite{wang2022privacy} employ collaborative inference in the frequency domain. In \cite{li2021deepobfuscator,xiao2020adversarial}, researchers used adversarial training to balance the deterioration of the reconstruction attack performance with the preservation of the face recognition performance. Nevertheless, methods in this category exhibit poor adaptability and generalizability, making them ineffective against adaptive attacks. To summarize, Table \ref{tab:comparision_work} offers a detailed comparison, highlighting the distinctions between the proposed method in this paper and the state-of-the-art techniques.

\section{Preliminaries}\label{section:Preliminaries} 

\subsection{Face Recognition Systems}
In this paper, we consider two typical architectures of face recognition systems, including uploading facial images and facial features as shown in Fig.\ref{fig:multics}.

\textit{Uploading facial images}: The current mainstream face recognition systems require users to upload facial images. The system architecture is shown in Fig.\ref{fig:multics}. B. Once the facial images are transmitted to a service provider, the service provider usually executes some preprocessing operations (such as basic image transformation and face alignment). After preprocessing, the facial images are input into a deep learning-based facial feature extractor to obtain facial feature vectors. Subsequently, similarity calculations are performed between the obtained feature vector and the facial feature vectors of other faces in face verification situations. The resulting similarity scores are compared with a threshold value to verify or confirm the identity of the face. This architecture usually risks privacy leakage when the service provider is untrustworthy. For example, after receiving users' facial images, the service provider may conduct a face attribute estimation attack and even stealthily train a deepfake detection model based on these facial images without the users' permission.

\textit{Uploading facial features}: In recent years, the user-server architecture was proposed to avoid privacy leakage caused by the original facial image transmission \cite{eshratifar2019jointdnn,ko2018edge}. The user-server architecture is shown in Fig.\ref{fig:multics}. A. In this architecture, the face recognition network is partitioned into a shallow model and a server model. They are deployed on the user side and service provider, respectively. In such a scenario, the user only needs to upload the feature vector $V$ of a facial image $X$ extracted by the shallow model $E_c(\cdot)$. The server model $E_s(\cdot)$ on the service provider subsequently recognizes the identity based on the received facial feature $V$. In this architecture, the service provider obtains facial features instead of facial images. It can avoid exposure the of the original facial images. However, recent researches show that adversaries can conduct a reconstruction attack to recover the original facial images by using pretrained face reconstruction models or surrogate models of $E_c(\cdot)$. That is to say, this strategy does not effectively prevent potential attackers from stealing users' private data.

\section{Threat Model And Evaluation Metrics}

In this section, we first present the threat model considered in this paper, and the evaluation metrics are also introduced. Note that both of the architectures discussed in Section \ref{section:Preliminaries} are taken into account from various levels.

\subsection{Threat Model}

In this paper, we consider a practical scenario where the service provider is untrustworthy and unreliable, with the potential risk of facial data exposure (because the service provider is usually controlled by a private company, and the private company may try to reconstruct the user's facial data or even exchange the data with others for benefits). Moreover, we consider the latent threats in two architectures, including uploading facial images and facial features. The goals and knowledge of the attacker are detailedly presented as follows.

\subsubsection{Case 1: Uploading Facial Features}

If one user submits his/her facial features to a service provider, the attacker (such as a service provider or other entities) can obtain the facial features. Because the facial features are computed based on the facial images and the shallow model of the face recognition service model, it is impractical for the attacker to use the facial features directly on other service models for privacy theft (because the parameters of different shallow models are different). Thus, it is necessary to reconstruct the facial image first before conducting data abuse or attribute estimation attacks. Therefore, we take into account two popular privacy leakage situations when uploading facial images, including data abuse and attribute estimation attacks. For uploading facial features, we only consider the reconstruction attacks.

\textbf{Reconstruction Attack}:
For the reconstruction attack, we consider two situations based on the differences of the attacker's knowledge. The knowledge and strategy of the attacker are presented as follows:

\textit{Level 1}: An attacker may have an auxiliary dataset that has a similar distribution to the training dataset\cite{cole2017synthesizing,dosovitskiy2016inverting,he2019model,mai2018reconstruction,zhmoginov2016inverting}. Then, the attacker can obtain many facial images and feature pairs by using the auxiliary dataset to query the shallow model. Leveraging this knowledge, the attacker can train a reconstruction model $R(\cdot)$ that can reconstruct the original facial images $X=\{x_1,x_2,x_2...x_N\}$ from the given facial features $Z=\{z_1,z_2,z_2...z_N\}$ (where $z_i=E_c(x_i)$ and $N$ represents the number of facial images collected by the attacker) by optimizing:
\begin{equation}
    \mathcal{L}_R(Z, X)=\sum_{i=1}^N\left\|x_i-R\left(z_i\right)\right\|_2
\end{equation}
After this, the attacker can obtain the reconstructed facial images $\hat{x_i}=R(z_i)$.

\textit{Level 2}: An attacker may obtain the shallow model $E_c(\cdot)$. This can be easily achieved. For instance, the attacker can purchase a client device from the face recognition service provider\cite{erdougan2022unsplit}. Therefore, given the facial features $z$, the attacker can directly minimize the distance of the facial features between the original images and reconstructed images:
\begin{equation}
\label{eq:op_recons}
\mathcal{J}(\hat{x},z) = \|E_c(\hat{x})-z\|_2 + \mathrm{TV}(\hat{x})
\end{equation}
where $\mathrm{TV}(\cdot)$ means the total variation of the reconstructed images as mentioned in \cite{erdougan2022unsplit}:
\begin{equation}
\label{eq:tv_loss}
\mathrm{TV}(x)=\sum_{i, j}\left(\left(x_{i, j-1}-x_{i, j}\right)^2+\left(x_{i+1, j}-x_{i, j}\right)^2\right)^{\frac{\beta}{2}}.
\end{equation}
By minimizing this loss function, the difference between neighboring pixels can be made as small as possible so that the optimized image has as little low-frequency semantic information as possible. The optimization process to update the reconstructed $\hat{x}$ can be formalized as:
\begin{equation}
\label{eq:op_update}
\hat{x}_{i+1} = \mathrm{Clip}_{(0,255)}(\hat{x}_{i} + \alpha\cdot\mathrm{sign}(\nabla_{\hat{x}_{i}}\mathcal{J}(\hat{x}_{i},z)))
\end{equation}
where $\mathrm{Clip}(\cdot)$ is used to constrain the pixel values of the reconstructed images within 0 and 255, and $\alpha$ means the step size of the iteration. 

Both of the aforementioned two levels are considered to conduct a reconstruction attack in this paper.

\subsubsection{Case 2: Uploading Facial Images}

If one user directly submits his/her original facial images to a service provider, an attacker (such as the service provider or other entities) may obtain the facial images. The attacker may further try to recover the original facial images based on the received (protected or obfuscated) facial images. In this case, the attacker is able to obtain the facial images, the corresponding facial feature pairs, and the shallow model. Then, he/she can also conduct a reconstruction attack as well, which is the same as that of \textit{Case 1} in Section 4.1.2. Hence, two potential risks, including data abuse and attribute estimation attacks, are considered here.

\textbf{Data Abuse}:
When an attacker obtains the original facial images, the attacker may stealthily utilize the facial images on other face recognition models without the data owner's permission\cite{Hill21}. Therefore, for this situation, we mainly validate whether the proposed method can achieve high utility on authorized face recognition models while performing poorly on others.

\textbf{Attribute Estimation Attack}:
When an attacker obtains the original facial images, he/she may try to analyze the user's private information (such as age, gender, mood, etc.) for advertising recommendation by conducting an attribute estimation attack\cite{han2017heterogeneous,karkkainen2021fairface} based on an attribute model $A(\cdot)$ and the given original facial images $X$:
\begin{equation}
\label{eq:train_aa}
    Q = \{A(x_i)|x_i \in X, i\in (0,N)\}
\end{equation}
where $Q$ denotes the face attribute vectors. 

In regard to all the threat models mentioned above, Table \ref{tab:threat_model} offers a more systematic exposition.

\begin{table}[]
\caption{Threat Model.}
\label{tab:threat_model}
\centering
\setlength{\tabcolsep}{1pt}
 \renewcommand{\arraystretch}{2}
\begin{tabular}{ccccc}
\toprule
\multicolumn{1}{l}{} & \multicolumn{2}{c}{Uploading Facial Features} & \multicolumn{2}{c}{Uploading Facial Images} \\ 
\cline{2-5} 
\multicolumn{1}{l}{} & \multicolumn{2}{c}{Reconstruction Attack}  & \makecell[c]{Data\\ Abuse}  & \makecell[c]{Attribute \\Attack} \\ 
\hline
Knowledge & \makecell[c]{facial images \\and feature pairs} & \makecell[c]{shallow \\model} & \makecell[c]{facial \\images} & \makecell[c]{facial \\images}  \\

Capability & \multicolumn{2}{c}{\makecell[c]{obtaining facial \\features}} & \multicolumn{2}{c}{\makecell[c]{obtaining facial \\images}} \\

Goal & \multicolumn{2}{c}{\makecell[c]{reconstructing \\original facial images}}&\makecell[c]{used in\\other models} & \makecell[c]{attribute \\estimation} \\ 
\bottomrule
\end{tabular}
\end{table}

\begin{figure*}
\centering
\includegraphics[width=13cm]{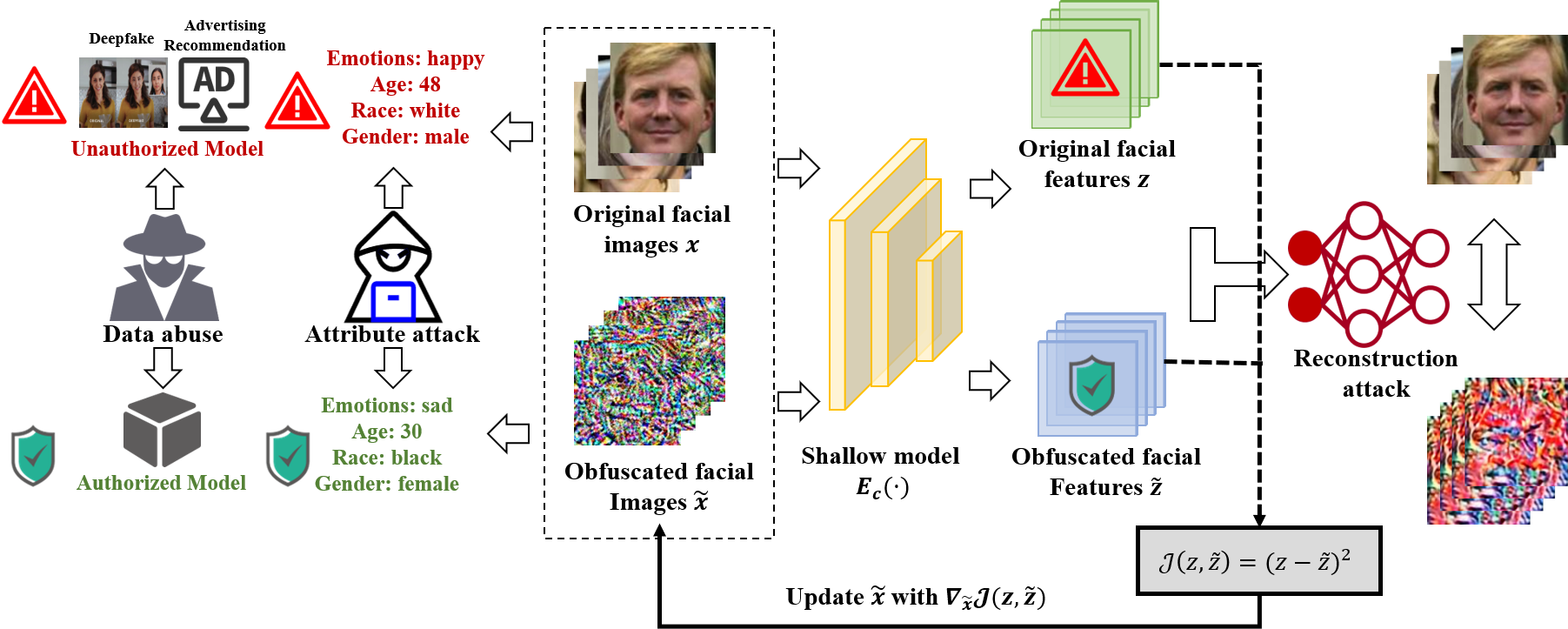}
\caption{The framework of PMT.}
\label{fig:framework}
\end{figure*}

\section{Design of PMT}

In this section, we first provide an overview of PMT, and then the detailed design of PMT is also presented. 

\subsection{Overview of PMT}
In this paper, we mainly consider two typical scenarios, including uploading facial images and facial features in face recognition applications. In order to defend against facial data abuse and privacy leakage, we mainly want to achieve two goals. One is to enable service providers to extract accurate features from the facial data to achieve face recognition tasks without violating users' privacy. The other one is that the user's data should exclusively serve the authorized face recognition model and restrict performance on other unauthorized models. In order to pursue the aforementioned goals, we propose a privacy minimization transformation dubbed PMT to filter out task-agnostic but private information. The overall framework is shown in Fig.\ref{fig:framework}. Our idea is to minimize the difference between features extracted from the original facial data and features derived from the obfuscated facial data using a shallow model. Then, considering that most face recognition services usually perform various preprocessing operations on the received data (such as adding noises), the perturbation enhancement technique is presented to improve the robustness of the proposed PMT. Finally, a multiple-restriction mechanism is proposed to authorize multiple service models simultaneously, which further improves the scalability of PMT. The design of PMT is detailedly presented as follows.

\subsection{Initial Formulation for Data Minimization}

Many methods have been proposed to defend against data abuse and privacy leakage in face recognition applications. However, most of the existing methods only consider uploading facial images or uploading facial features. Besides, most of them also need to acquire the attacker's strategy or complete face recognition model, which is unrealistic in practical application. Therefore, we propose a plug-and-play defense method named PMT for the user side without the requirement for a complete face recognition model and knowledge of attack strategy. In this section, we present the initial version of PMT. 


We take inspiration from past works that utilized adversarial examples for privacy-preserving. In the case of neighboring data under p-norm metrics, the features extracted by deep neural networks may display substantial variations, which can be formalized as:
\begin{equation}
\label{eq:adv}
\begin{aligned}
    \max \|f(x_{adv})-f(x)\|_2 \\
    \text{subject\ to \ } \|x_{adv}-x\|_p\leq\epsilon
\end{aligned}
\end{equation}
where $x_{adv}$ denotes the adversarial example and $\epsilon$ is the budget of adversarial perturbations. Adversarial attacks aim to deceive deep learning models to output false results by generating adversarial samples that are visually close to the original facial data. In contrast, in this paper, our goal is that the outputs of the original facial images and the obfuscated facial images after inputting the face recognition model are as consistent as possible. Moreover, the original facial images and the obfuscated facial images should be visually different, thereby preventing manual vision privacy theft (e.g., people extract facial features visually). Besides, we also try to ensure that the obfuscated facial images can maintain satisfactory performance on authorized models and restrict the performance on other unauthorized models. Hence, we proposed the PMT to achieve above goals. The PMT contains the following steps. First, we start by randomly initializing the obfuscated facial images. Then, we subsequently perform forward computations with the user's shallow model on the initialized obfuscated facial images and the user's original facial images respectively. Finally, we try to minimize the distance between the original facial images $x$ and obfuscated facial images $\widetilde{x}$ output from the shallow model:
\begin{equation}
\label{eq:pmtv1}
\begin{aligned}
    \min \|f(\widetilde{x})-f(x)\|_2 \\
\end{aligned}
\end{equation}
By applying backpropagation, we derive the gradient sign to update the initialized obfuscated facial images. The objective of each update process is to maximize the utility of information. Algorithm \ref{agm:pmt} shows the complete algorithm of our PMT.

\begin{algorithm}
    \caption{Privacy Minimization Transformation}
    \label{agm:pmt}
    \begin{algorithmic}[1]  
        \Require The shallow model $E_{c}(\cdot)$; facial images $x$; a maximum of iterations $K$; update step $\alpha$
        \Ensure  obfuscated facial images $\widetilde{x}$; obfuscated facial features $\widetilde{z}$. \\
        $\widetilde{x} \leftarrow \mathrm{randn\_like}(x)$ \\
        // Initialize the obfuscated facial images $\widetilde{x}$\\
        $z\leftarrow E_{c}(x)$
	\For {$i=1 \to K$}
            \State $\widetilde{z}\leftarrow E_{c}(\widetilde{x}_{i})$
            \State$\mathcal{J}(z,\widetilde{z})\leftarrow (z-\widetilde{z})^{2}$
            \State$\widetilde{x}_{i+1} \leftarrow \widetilde{x}_{i} + \alpha \cdot\nabla_{\widetilde{x}_{i}}\mathcal{J}(z,\widetilde{z})$
            \State// Update the obfuscated facial images $\widetilde{x}$
	\EndFor
        \State\Return $\widetilde{x}_{K},\widetilde{z}$
	\end{algorithmic}
\end{algorithm}

\subsection{Improve Robustness with Enhanced Perturbation}
We have constructed a basic version of PMT as above Section 5.2. However, in a real-world scenario, before inputting facial data into the service provider's model, a standardized service process performs various preprocessing operations on the received data, during which most of the privacy-preserving methods will fail to defend against privacy thieves or have a significant loss of accuracy. Additionally, prior investigations have demonstrated that the privacy-preserving techniques utilizing adversarial perturbations are highly susceptible to vulnerabilities\cite{xu2017featuresqueezing}, and even commonly used data transformations can exert substantial effects on this kind of method, limiting their practicality in real-world scenarios\cite{deb2023faceguard}. In this section, we present a method to strengthen the robustness of our PMT without reducing data availability. In other words, we need to design a robust enhancement method to optimize the obfuscated facial images so that the corresponding loss values are not sensitive to unknown noises or transformations. To achieve this goal, we propose a robustness enhancement method that contains three phases as follows. Algorithm \ref{agm:pm} shows the complete algorithm of the proposed robustness enhancement method.

\textbf{Initialization for PMT}. Note that there is no requirement for facial data structural similarity. We have mentioned that our modification on features differs from the generation of adversarial examples, which aims to fool humans visually. The only goal is to preserve the utility of our data while preventing privacy leakage. Therefore, we initialize the obfuscated facial images randomly, where three versions of initialization are adopted, including Gaussian noise, random substitution, and Gaussian blur. In this way, the initial facial images are completely different from the user's original facial images, thereby protecting the privacy of the users' facial images from leaking. Then, the initial obfuscated facial images is updated based on the output of the shallow model to enhance the utility of the processed data.

\textbf{Data augmentation and layer aggregation for PMT}. After finishing data initialization, data augmentation and layer aggregation are also employed to enhance the robustness of PMT. In this paper, we mainly consider several commonly used preprocessing transformations for service providers, including random affine, random noise, and their combination. We use data augmentation $\mathrm{T}(\cdot)$ to ensure the feature consistency in our shallow model $E_{c}(\cdot)$ between $E_{c}(\mathrm{T}(x))$ and $E_{c}(\mathrm{T}(\widetilde{x}))$. Besides, we learn from the intermediate-level attack introduced by Huang et al.\cite{huang2019enhancing} to enhance the effectiveness of perturbation. Specifically, we achieve this by concatenating (concat function) the output of multi-layer features. On the whole, before each optimization iteration, we first perform data augmentation and then use multi-layer representations of shallow models as features of facial images for optimization so that the original facial data $x$ features and the obfuscated facial data $\mathrm{PMT}(x)$ features are aligned. For the selection of different augmentation strategies, we conduct experiments to figure it out in the Section 6.

\textbf{Translation-invariant perturbation for PMT}. After completing the above steps, a translation-invariant perturbation mechanism is proposed. Previous works\cite{dong2019evading, wang2021feature, lin2019nesterov,wei2023physically} have proved that the variance of the adversarial perturbations can boost the transferability and robustness of adversarial examples theoretically and empirically. Similarly, we follow the technique of performing convolution operation on the gradients calculated after backpropagation\cite{dong2019evading} while optimizing the perturbations to reduce the variance. It means that the adjacent pixels of the perturbations should be the same as possible. Formally, we define our convolution as: $\textbf{W}\otimes\delta$, where $\textbf{W}$ is the convolution kernel and $\otimes$ stands for "same" convolution. For selections of the kernel, we have:
\begin{enumerate}
    \item A linear kernel that $\hat{W}_{i, j}=(1-|i| / (k+1)) \cdot(1-|j| / (k+1))$, and $\textbf{W}_{i, j}=\hat{W}_{i, j} / \sum_{i, j} \hat{W}_{i, j}$;
    \item A Gaussian kernel that $\hat{W}_{i, j}=\frac{1}{2 \pi \sigma^2} \exp \left(-\frac{i^2+j^2}{2 \sigma^2}\right)$ where $\sigma = k/\sqrt{3}$ is the standard deviation to make the radius of the kernel be $3 \sigma$, and $\textbf{W}_{i, j}=\hat{W}_{i, j} / \sum_{i, j} \hat{W}_{i, j}$.
\end{enumerate}
In light of the strong association between the choice of kernels with different sizes and types, and the practical and privacy considerations of the data, we will make a comparison by empirical experiments in Section 6. Relevant experimental results show that the robustness of PMT has been further improved after employing the above steps. It can be better applied to real-world face recognition systems.

\begin{algorithm}
    \caption{Perturbation Enhancement}
    \label{agm:pm}
    \begin{algorithmic}[1]  
        \Require The shallow model $E_{c}(\cdot)$; facial images $x$; obfuscated facial images $\widetilde{x}$; convolution kernel $\textbf{W}$; data augmentation function $\mathrm{T}(\cdot)$
        \Ensure enhanced perturbations $\delta$ \\
        $z,\widetilde{z}\leftarrow[\ ],[\ ]$
        \For {$l=1 \to L$}
            \State $z\leftarrow \mathrm{Concat}(z,\mathrm{AvgPool2d}(E_{c}^{l}(\mathrm{T}(x))))$
            \State $\widetilde{z}\leftarrow \mathrm{Concat}(\widetilde{z},\mathrm{AvgPool2d}(E_{c}^{l}(\mathrm{T}(\widetilde{x}))))$
            \State// Calculate the multi-layer features $z$ and $\widetilde{z}$
        \EndFor
        \State $\mathcal{J}(z,\widetilde{z})\leftarrow (z-\widetilde{z})^{2}$
        \State $\delta \leftarrow \textbf{W}\otimes\frac{\nabla_{\widetilde{x}_{i}}\mathcal{J}(z,\widetilde{z})}{\|\nabla_{\widetilde{x}_{i}}\mathcal{J}(z,\widetilde{z})\|_{1}}$
        \State// Calculate the convolution of the perturbations
        \State \Return $\delta$
	\end{algorithmic}
\end{algorithm}

\subsection{Improve Scalability with Multiple-Restriction}
Notably, users may simultaneously authorize one facial image to multiple service models. However, previous studies have typically concentrated on generating perturbations tailored to specific tasks to safeguard data privacy, implying the need for different perturbations for different service models, which is often unfeasible and resource-wasting. Consequently, generating universal data perturbations that can be employed across multiple authorized models becomes paramount. Based on the proposed PMT, we further authorize multiple service models by setting multiple-restriction as follows:  
\begin{equation}
\label{eq:pmtv2}
\begin{aligned}
    \min \sum_{i=1}^{M}\|f_{i}(\widetilde{x})-f_{i}(x)\|_2 \\
\end{aligned}
\end{equation}
where $f_{i}\in\{f_{1},f_{2},f_{3}...f_{M}\}$ and $M$ is the number of the authorized service models. By applying multiple-restriction to PMT, we obtain a scalable PMT named MR-PMT. To validate the effectiveness of MR-PMT, we conducted comprehensive experiments with various models in Section 6.
\section{Experiment}
In this section, we comprehensively evaluate the performance of the proposed PMT. 

\subsection{Experimental Settings}

\textbf{Device Configuration}: All experiments were conducted on a server equipped with Intel i9-9900K, 3.60 GHz processor, 32GB RAM, NVIDIA GeForce RTX 2080TI. PyCharm, PyTorch, and RobFR\cite{RobFR} are used to deploy the model and complete other relevant experiments.

\textbf{Datasets}: We use the following datasets in our experiments for evaluation.
\begin{itemize}
    \item LFW\cite{2008Labeled} contains 13K labeled facial images in the wild from 5.7K different identities and 6K face pairs.

    \item AgeDB-30\cite{2017AgeDB} contains more than 12K images from 570 identities and 6K face pairs for evaluation.

    \item CFP-FF\cite{2016Frontal} contains 5K images from 500 identities and 7K face pairs for evaluation.

    \item CALFW\cite{2017Cross} contains 12k cross-age facial images from 4k different identities.

\end{itemize}
All the facial images of the four datasets were cropped to 112×112 with multitask convolutional neural network (MTCNN)\cite{zhang2016joint}.

\textbf{Face Recognition Models:} 
We use the following models: MobileFace\cite{chen2018mobilefacenets}, IR50-ArcFace\cite{deng2019arcface}, IR50-SphereFace\cite{liu2017sphereface}, IR50-Softmax, MobileNet\cite{howard2017mobilenets} and ResNet50\cite{he2016deep} trained on CASIA-WebFace\cite{yi2014learning} dataset for 100 epochs with different loss functions. Following that, the first five layers of the models mentioned above are separated to function as the shallow model shared with the registered users, while the service provider holds the whole model. Hence, users can upload facial images or features to enjoy face recognition services.

\textbf{Face Reconstruction Models:} Following the configurations mentioned in \cite{wang2023privacy}, we modified the Unet\cite{ronneberger2015u} architecture to construct a reconstruction network named RecFace. We first generated the facial pairs using MobileFace and trained RecFace on LFW. 

\begin{table}[]
\centering
\caption{PMT performance under different initialization statuses.}
\label{tab:initial_mode}
\renewcommand{\arraystretch}{1.2}
\begin{tabular}{cccc}
\toprule
\multicolumn{2}{c}{\multirow{2}{*}{}} & Authorized$\uparrow$ & Unauthorized$\downarrow$ \\ 
\hline
\multirow{3}{*}{Initial Status} 
& Random Sort & 99.0\% & 87.5\% \\
& Random Blur & 98.1\% & 99.0\% \\
& Random Noise & 99.0\% & 10.6\% \\
\bottomrule
\end{tabular}
\end{table}

\begin{table}[]
\centering
\setlength{\tabcolsep}{1mm}{
\caption{PMT performance under different augmentation modes and layer aggregation.}
\label{tab:augment_type}
\renewcommand{\arraystretch}{1.2}
\begin{tabular}{cccc}
\toprule
\multicolumn{2}{c}{\multirow{2}{*}{}} & Authorized$\uparrow$ & Unauthorized$\downarrow$ \\ 
\hline
\multirow{4}{*}{Augmentation} 
& Mix & 100\% & 19.2\% \\
& None & 100\% & 15.8\% \\
& Random Noise & 99.1\% & 8.7\% \\
& Random Affine & 99.0\% & 44.2\% \\
\hline
\multirow{2}{*}{Layer Aggregation} 
& Yes & 99.0\% & 12.5\% \\
& No & 99.0\% & 15.4\% \\ 
 \bottomrule
\end{tabular}}
\end{table}

\begin{table}[]
\centering
\caption{PMT performance under different kernel types.}
\label{tab:kernel_type}
\renewcommand{\arraystretch}{1.2}
\begin{tabular}{cccc}
\toprule
\multicolumn{2}{c}{\multirow{2}{*}{}} & Authorized$\uparrow$ & Unauthorized$\downarrow$ \\ 
\hline
\multirow{3}{*}{Kernel} & None & 98.1\% & 14.4\% \\
& Linear Kernel & 98.1\% & 19.2\% \\ 
& Gaussian Kernel & 99.0\% & 9.6\% \\
 \bottomrule
\end{tabular}
\end{table}

\begin{figure}[]
\centering
\subfloat{\includegraphics[width=3in]{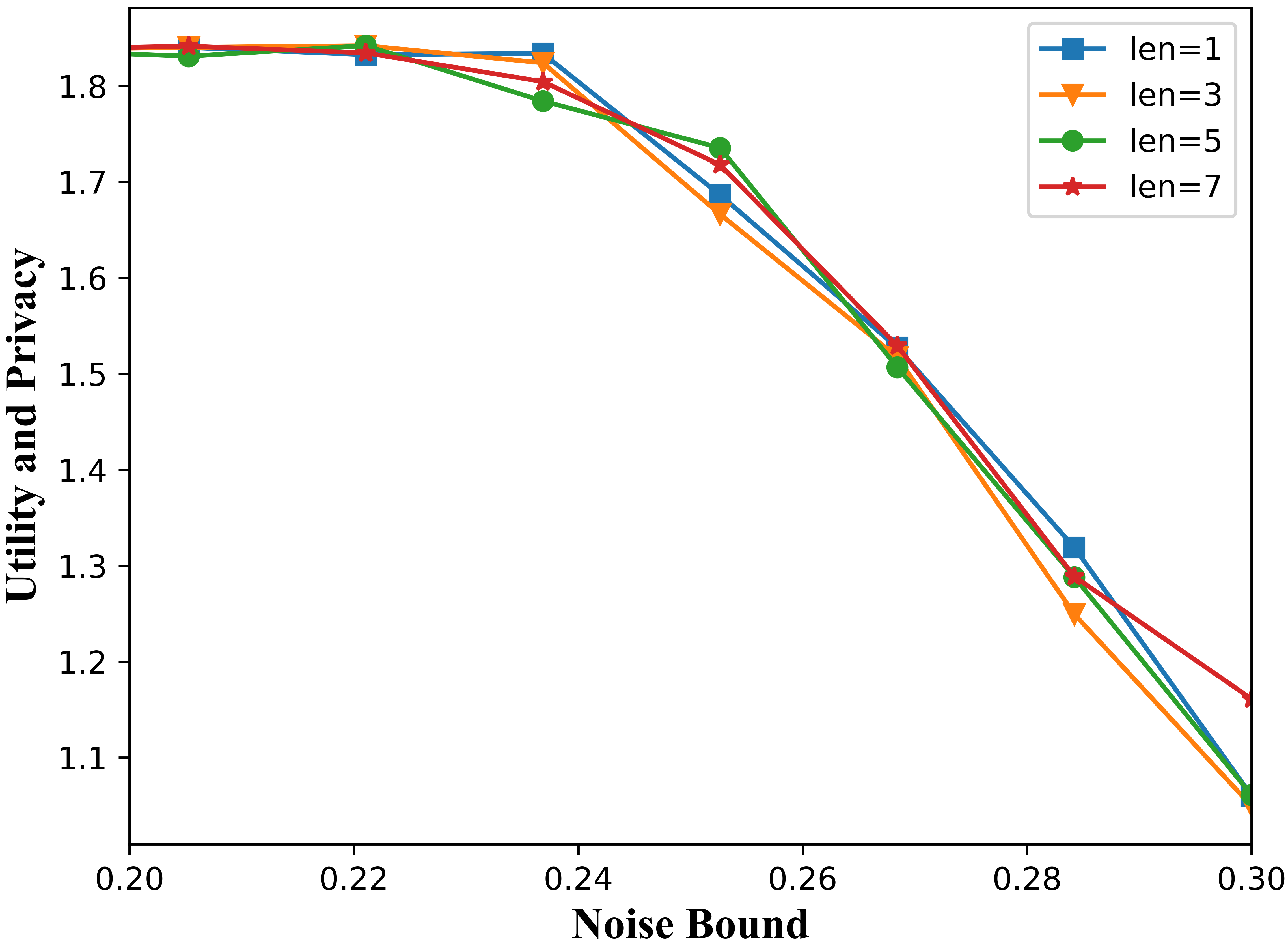}}
\caption{
\justifying
$UP(\delta, \kappa)$ of PMT with different noise bounds $\delta$ and kernel length on AgeDB30. 
}
\label{fig:kernel_length}
\end{figure}

\begin{figure}[]
\centering
\subfloat{\includegraphics[width=3in]{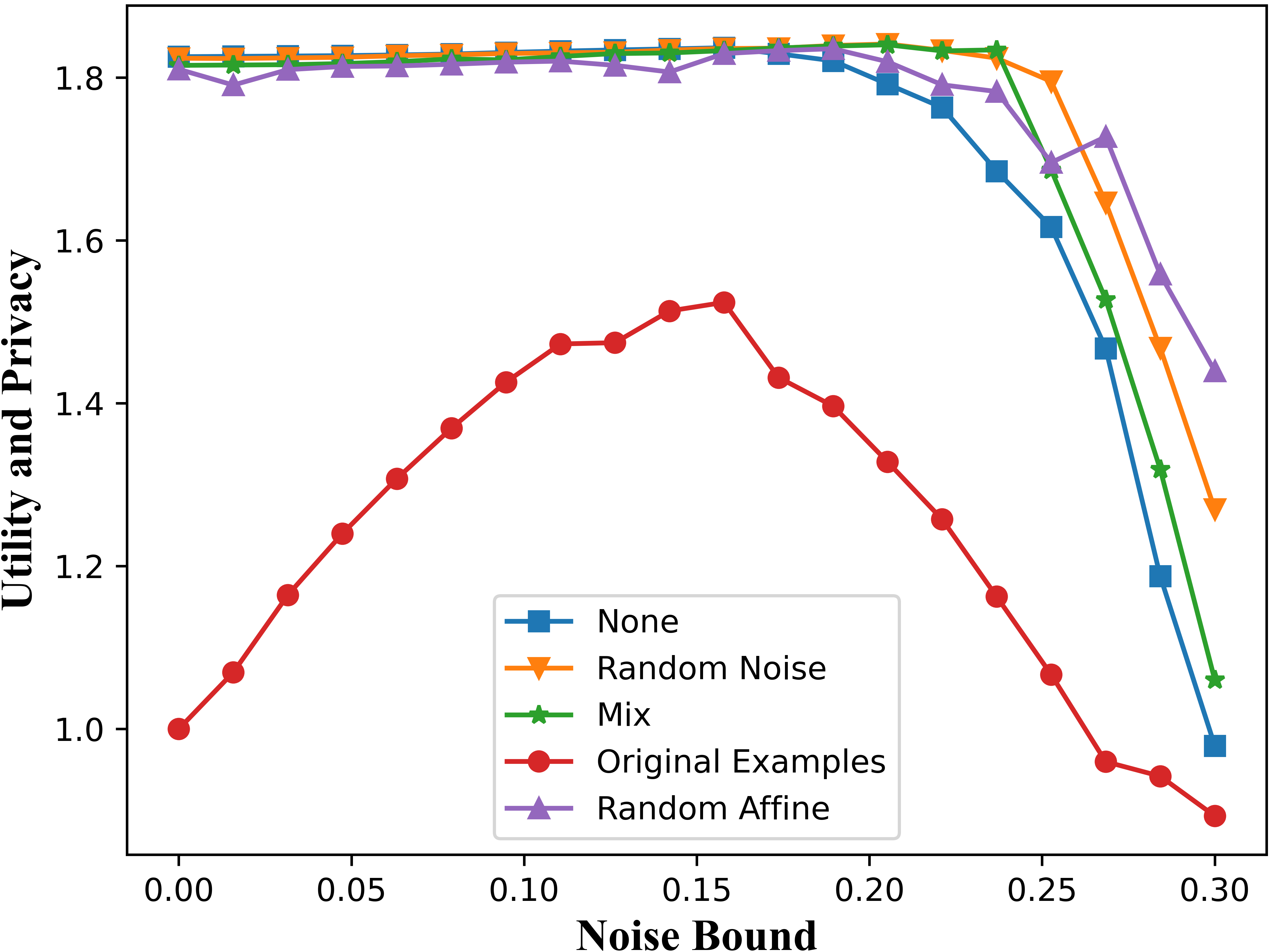}}
\caption{
\justifying
$UP(\delta, \kappa)$ of PMT with different noise bounds $\delta$ and augmentation modes for facial images on AgeDB30. 
}
\label{fig:plot_augmode}
\end{figure}

\textbf{Face Attribute Models:} For face attribute estimation, we choose the popular DeepFace\cite{deepface} package, including age, gender, race, and emotion classifiers. Given the images, we followed the steps of detecting and cropping the existing faces using the face detector backend. The function returns a list of attribute dictionaries for each face appearing in the image. For attackers with access to the facial features, we used face reconstruction models to obtain images and then conducted an attribute estimation attack.

\textbf{Parameters configuration of PMT:} Note that we have some alternative parameters to be set before our experiments. In order to obtain better parameter combinations, we conducted experiments on the AgeDB30 dataset to compare the performance impact of different parameter selections on authorized and unauthorized models (Authorized: IR50-Softmax, Unauthorized: MobileFace). For different selections of initial status, we consider three statuses. Experiment results shown in Table \ref{tab:initial_mode} indicate that using ``random noise'' outperforms others. Moreover, we adopt different augmentation methods and layer aggregation to enhance the robustness of obfuscated facial data. From Table \ref{tab:augment_type}, we can see that employing ``random noise'' and layer aggregation can achieve much lower accuracy on unauthorized models and acceptable degradation on the accuracy of authorized models. Specifically, for the robustness of these augmentation methods toward different magnitudes of noises, the experiment results shown in Fig.\ref{fig:plot_augmode} verify the superiority of ``random noise''. For the selection of different kernel types and length, the results shown in Table \ref{tab:kernel_type} and Fig.\ref{fig:kernel_length} demonstrate that ``Gaussian kernel'' outperform others with length set to $1$. Therefore, in the following experiments, we have confirmed the selection of the basic parameters of PMT: using random noise for initialization, using random noise for augmentation, employing layer aggregation, and using Gaussian kernel for perturbation convolution (length is set to 1). 

\begin{table*}[]
\centering
\caption{Experimental results of white-box and black-box reconstruction attacks in terms of SSIM, PSNR, COS, and SRRA.}
\label{tab:defense_recons}
\renewcommand{\arraystretch}{1.2}
\resizebox{1.5\columnwidth}{!}{
\begin{tabular}{ccccllllll}
\toprule
\multirow{2}{*}{Dataset} & \multirow{2}{*}{Method} & \multicolumn{4}{c}{White-Box} & \multicolumn{4}{c}{Black-Box} \\ 
\cline{3-10} 
&  & SSIM$\downarrow$ & PSNR$\downarrow$ & COS$\uparrow$ & SRRA$\uparrow$ & SSIM$\downarrow$ & PSNR$\downarrow$ & COS$\uparrow$ & SRRA$\uparrow$ \\
\hline
\multirow{2}{*}{LFW} 
& Unprotected & 0.62 & 14.61 & 0.96 & 1.00 & 0.99 & 34.40 & 0.98 & 1.00 \\
& Random Noise & 0.38 & 12.56 & 0.69 & 1.00 & 0.78 & 15.55 & 0.55 & 1.00  \\
& Ours & 0.03 & 11.31 & 0.83 & 1.00 & 0.03 & 11.21 & 0.71 & 1.00 \\ 
\hline
\multirow{2}{*}{AgeDB-30} 
& Unprotected & 0.58 & 14.34 & 0.96 & 1.00 & 0.99 & 30.05 & 0.96 & 1.00 \\
& Random Noise & 0.33 & 12.31 & 0.69 & 1.00 & 0.71 & 15.81 & 0.54 & 1.00 \\
& Ours & 0.02 & 11.08 & 0.81 & 1.00 & 0.02 & 11.07 & 0.70 & 1.00 \\ 
\hline
\multirow{2}{*}{CALFW} 
& Unprotected & 0.62 & 15.14 & 0.97 & 1.00 & 0.99 & 31.71 & 0.96 & 1.00 \\
& Random Noise & 0.38 & 12.98 & 0.72 & 1.00 & 0.80 & 17.01 & 0.57 & 1.00 \\
& Ours & 0.02 & 11.60 & 0.83 & 1.00 & 0.03 & 11.57 & 0.72 & 1.00 \\ 
\hline
\multirow{2}{*}{CFP-FF} 
& Unprotected & 0.53 & 11.83 & 0.94 & 1.00 & 0.98 & 29.05 & 0.95 & 1.00 \\
& Random Noise & 0.32 & 10.32 & 0.67 & 1.00 & 0.61 & 12.23 & 0.46 & 1.00 \\
& Ours & 0.02 & 9.32 & 0.81 & 1.00 & 0.03 & 9.17 & 0.68 & 1.00 \\
\bottomrule
\end{tabular}}
\end{table*}

\subsection{Evaluation Metrics}

To evaluate the effectiveness of the proposed PMT, we present the evaluation metrics for defending against the above attacks.

\textbf{Privacy-Preserving and Data Utility Evaluation Metrics}: We employ Structural SIMilarity (SSIM) \cite{wang2004image}, Peak Signal-to-Noise Ratio (PSNR) \cite{wang2023privacy} to evaluate the performance of defending against the reconstruction attack (namely the effect of privacy-preserving). SSIM and PSNR are used to measure the structural similarity, i.e., whether the reconstructed facial image and the original facial image are similar from human vision. Higher SSIM and PSNR indicate that the reconstructed image and the original image are more visually similar. The Cosine similarity (COS) \cite{wang2018cosface} of facial features and the success rate of reconstruction attack (SRRA) are used to evaluate the reconstructed facial image utility. COS is used to measure whether the features of the reconstructed facial image and that of the original facial image are similar. SRRA is employed to evaluate whether the reconstructed image is recognized as the original image by the face recognition model. Note that our goal is to prevent the original data from being reconstructed while ensuring the utility of the obfuscated data or even reconstructed data. Hence, the objective of this paper is to maintain high COS and SRRA while reducing SSIM and PSNR.

Additionally, we also try to find a satisfying trade-off of utility-privacy. Therefore, to evaluate the trade-off between privacy-preserving and data utility, we define the $UP(\sigma,\kappa)$ which can be expressed as: 
\begin{equation}
\label{eq:up}
\begin{aligned}
    UP(\sigma,\kappa) = & \mathop{\mathbb{E}}_{
    x \sim \mathcal{N}(X,\sigma^2)
    \atop
    \widetilde{x} \sim \mathcal{N}(\widetilde{X},\sigma^2)
    }\{\mathbb{I}(\mathrm{cos}(f(\widetilde{x}),f(x))>\kappa) \\
    & + \frac{1}{K}\sum_{i=0}^{K}\mathrm{Dist}(\mathrm{m}_{i}(x),\mathrm{m}_{i}(\widetilde{x}))\}
\end{aligned}
\end{equation}

The utility-privacy measurement $UP(\sigma,\kappa)$ considers both the main task performance and the privacy-preserving where $\sigma$ and $\kappa$ stand for the magnitude of environmental noise in the real world and the threshold of face similarity, respectively. The first part of the Equation \ref{eq:up} represents the impact of a privacy protection method on the performance of the target task. Note that $\widetilde{X}$ represents the ${X}$ processed by PMT (named $PMT(X)$, i.e. the obfuscated data). $\mathbb{I}(\cdot)$ denotes the identity function, which is used to compute the similarity of two facial images. The second part of the Equation \ref{eq:up} represents the degree of protection for the privacy information measured by various privacy-infringing models. Note that $m_i$ represents various privacy-infringing models (e.g., attribute estimation model). $m_i(x)$ represents the privacy information extracted from $x$. $\mathrm{Dist}(\cdot)$ denotes the similarity measurement of the extracted privacy information from the original facial images and the obfuscated facial images. In this paper, in order to measure the degree of privacy protection easily, we use the LPIPS metric \cite{zhang2018unreasonable} to substitute the second part of the Equation \ref{eq:up}, thereby measuring the consistency of the original data and the obfuscated one (i.e., VGG serves as ``privacy-infringing model $m_i(\cdot)$'').

\textbf{Defense Data Abuse Evaluation Metrics}: To evaluate data availability on a specific face recognition model, existing research usually employs face recognition accuracy. The existing definition of the face recognition accuracy only considers the distance between the obfuscated image $\widetilde{x_i}$ and the original image $x_i$. In this paper, to demonstrate the effectiveness of PMT, we define a more strict face recognition accuracy to evaluate. The presented accuracy not only considers the distance between the obfuscated image $\widetilde{x_i}$ and its original one $x_i$ but also takes into account the distance between the obfuscated image $\widetilde{x_i}$ and image $x_j$ with another identity. We define the proposed accuracy $acc$ as: 
\begin{equation}
\label{eq:fr_acc}
\begin{aligned}
acc = \mathop{\mathbb{E}}_{(x_{1},x_{2},y)\in\mathcal{D}_{test}}\{
    &\mathbb{I}(\mathrm{cos}(f(\widetilde{x}_{1}),f(x_{1}))>\kappa)\ and\ \\
    \mathbb{I}(&\mathrm{cos}(f(\widetilde{x}_{1}),f(x_{2}))>\kappa) = y)\} \\
\end{aligned}
\end{equation}
where $\mathcal{D}_{test}$ is the test set of the face recognition model, $y=\{0,1\}$ denotes whether $x_{1}$ and $x_{2}$ are the same identity, and $\kappa$ denotes the threshold of facial feature similarity (we usually set $0.2$ for this threshold). In simple terms, we measure the recognition accuracy $acc$ of authorized and unauthorized service models, and our goal is to gain high accuracy on authorized service models while performing poorly on unauthorized ones. That is, we employ $acc$ to measure the effectiveness of data abuse protection.

\textbf{Defense Attribute Estimation Attack Evaluation Metrics}: Given a facial image $x$, the potential attacker can obtain $A(x)$ using attribution model $A(\cdot)$. $A(x)$ is the face attribute vector. To evaluate the performance of defending against face attribute estimation attacks (using the absolute difference for age measurement), we use cosine similarity (called $CosSim$) between the obfuscated image attribute vectors and corresponding original image attribute vectors. $CosSim$ is defined as:

\begin{equation}
\label{eq:fr_acc}
\begin{aligned}
CosSim = \mathrm{cos}(A(\widetilde{x}),A(x))\\
\end{aligned}
\end{equation}
where $\widetilde{x}$ and $x$ denote the obfuscated image and corresponding original one, respectively. $A(\cdot)$ represents the attribution model. Therefore, $CosSim$ is used to evaluate the performance of PMT against attribute estimation attacks. The lower the value of $CosSim$, the better our defense method works.

\subsection{Defense against Face Reconstruction Attack}
We conduct reconstruction attacks under white-box and black-box settings on the authorized model (MobileFace). Table \ref{tab:defense_recons} summarizes the average values of SSIM, PSNR, and COS of the reconstructed images and original images. Note that the images are reconstructed respectively based on original features, processed features by PMT, and processed features by adding random noise with the same amplitude as the PTM. Different from the generation of adversarial examples, we should maintain high COS and SRRA while reducing SSIM and PSNR. 

\subsubsection{Level 1}
As listed in Table \ref{tab:threat_model}, we consider the real-world black-box condition where the attacker is only accessible to facial images and features pairs \cite{cole2017synthesizing,dosovitskiy2016inverting,he2019model,mai2018reconstruction,zhmoginov2016inverting}. Then, attackers use face reconstruction models to map the facial features to images. The experimental results are shown in Table \ref{tab:defense_recons}. Let's take the result on LFW as an example: compared with unprotected samples, SSIM drops from 0.99 to 0.03, PSNR drops from 34.40 to 11.21 while SRRA maintains high accuracy. This illustrates that our PMT can disrupt spatial information while maintaining the accuracy on authorized face recognition services. Fig.\ref{fig:defense_recons_vis} is the visualization of reconstructed images, where we can see that unprotected facial images have much semantic information after reconstruction, while the images processed with PMT remain confusing. PMT cannot only mislead unauthorized models but also humans since even humans cannot extract private information from the processed facial images and reconstructed ones. 

\subsubsection{Level 2}
\begin{figure*}[htbp]
\centering
\subfloat{\includegraphics[width=0.7in]{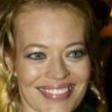}}
\hspace{0mm}
\subfloat{\includegraphics[width=0.7in]{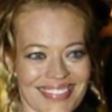}}
\hspace{0mm}
\subfloat{\includegraphics[width=0.7in]{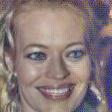}}
\hspace{0mm}
\subfloat{\includegraphics[width=0.7in]{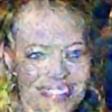}}
\hspace{0mm}
\subfloat{\includegraphics[width=0.7in]{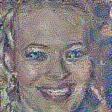}}
\hspace{0mm}
\subfloat{\includegraphics[width=0.7in]{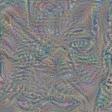}}
\hspace{0mm}
\subfloat{\includegraphics[width=0.7in]{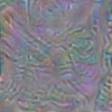}}
\hspace{0mm}
\subfloat{\includegraphics[width=0.7in]{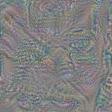}}

\vspace{-2.5mm}
\subfloat{\includegraphics[width=0.7in]{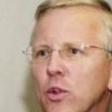}}
\hspace{0mm}
\subfloat{\includegraphics[width=0.7in]{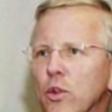}}
\hspace{0mm}
\subfloat{\includegraphics[width=0.7in]{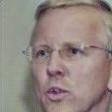}}
\hspace{0mm}
\subfloat{\includegraphics[width=0.7in]{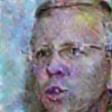}}
\hspace{0mm}
\subfloat{\includegraphics[width=0.7in]{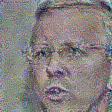}}
\hspace{0mm}
\subfloat{\includegraphics[width=0.7in]{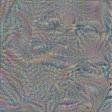}}
\hspace{0mm}
\subfloat{\includegraphics[width=0.7in]{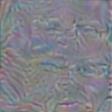}}
\hspace{0mm}
\subfloat{\includegraphics[width=0.7in]{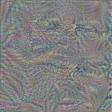}}

\vspace{-2.5mm}
\subfloat{\includegraphics[width=0.7in]{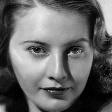}}
\hspace{0mm}
\subfloat{\includegraphics[width=0.7in]{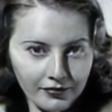}}
\hspace{0mm}
\subfloat{\includegraphics[width=0.7in]{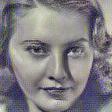}}
\hspace{0mm}
\subfloat{\includegraphics[width=0.7in]{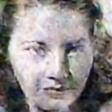}}
\hspace{0mm}
\subfloat{\includegraphics[width=0.7in]{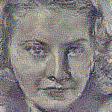}}
\hspace{0mm}
\subfloat{\includegraphics[width=0.7in]{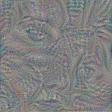}}
\hspace{0mm}
\subfloat{\includegraphics[width=0.7in]{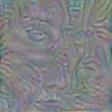}}
\hspace{0mm}
\subfloat{\includegraphics[width=0.7in]{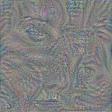}}

\vspace{-2.5mm}
\subfloat{\includegraphics[width=0.7in]{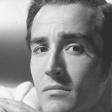}}
\hspace{0mm}
\subfloat{\includegraphics[width=0.7in]{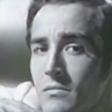}}
\hspace{0mm}
\subfloat{\includegraphics[width=0.7in]{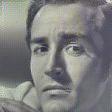}}
\hspace{0mm}
\subfloat{\includegraphics[width=0.7in]{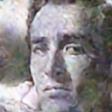}}
\hspace{0mm}
\subfloat{\includegraphics[width=0.7in]{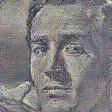}}
\hspace{0mm}
\subfloat{\includegraphics[width=0.7in]{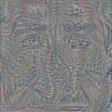}}
\hspace{0mm}
\subfloat{\includegraphics[width=0.7in]{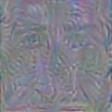}}
\hspace{0mm}
\subfloat{\includegraphics[width=0.7in]{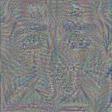}}

\vspace{-2.5mm}
\subfloat{\includegraphics[width=0.7in]{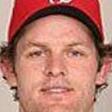}}
\hspace{0mm}
\subfloat{\includegraphics[width=0.7in]{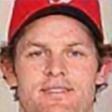}}
\hspace{0mm}
\subfloat{\includegraphics[width=0.7in]{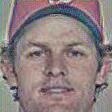}}
\hspace{0mm}
\subfloat{\includegraphics[width=0.7in]{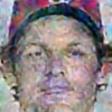}}
\hspace{0mm}
\subfloat{\includegraphics[width=0.7in]{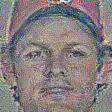}}
\hspace{0mm}
\subfloat{\includegraphics[width=0.7in]{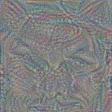}}
\hspace{0mm}
\subfloat{\includegraphics[width=0.7in]{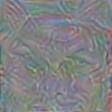}}
\hspace{0mm}
\subfloat{\includegraphics[width=0.7in]{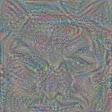}}

\vspace{-2.5mm}
\subfloat{\includegraphics[width=0.7in]{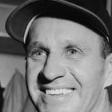}}
\hspace{0mm}
\subfloat{\includegraphics[width=0.7in]{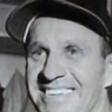}}
\hspace{0mm}
\subfloat{\includegraphics[width=0.7in]{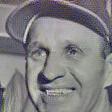}}
\hspace{0mm}
\subfloat{\includegraphics[width=0.7in]{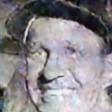}}
\hspace{0mm}
\subfloat{\includegraphics[width=0.7in]{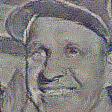}}
\hspace{0mm}
\subfloat{\includegraphics[width=0.7in]{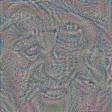}}
\hspace{0mm}
\subfloat{\includegraphics[width=0.7in]{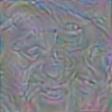}}
\hspace{0mm}
\subfloat{\includegraphics[width=0.7in]{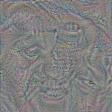}}

\vspace{-2.5mm}
\subfloat{\includegraphics[width=0.7in]{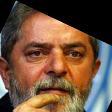}}
\hspace{0mm}
\subfloat{\includegraphics[width=0.7in]{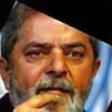}}
\hspace{0mm}
\subfloat{\includegraphics[width=0.7in]{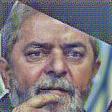}}
\hspace{0mm}
\subfloat{\includegraphics[width=0.7in]{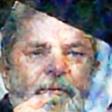}}
\hspace{0mm}
\subfloat{\includegraphics[width=0.7in]{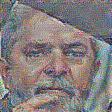}}
\hspace{0mm}
\subfloat{\includegraphics[width=0.7in]{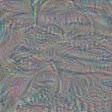}}
\hspace{0mm}
\subfloat{\includegraphics[width=0.7in]{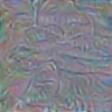}}
\hspace{0mm}
\subfloat{\includegraphics[width=0.7in]{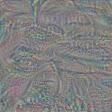}}

\vspace{-2.5mm}
\setcounter{subfigure}{0}
\subfloat[]{\includegraphics[width=0.7in]{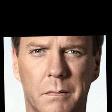}}
\hspace{0mm}
\subfloat[]{\includegraphics[width=0.7in]{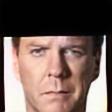}}
\hspace{0mm}
\subfloat[]{\includegraphics[width=0.7in]{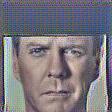}}
\hspace{0mm}
\subfloat[]{\includegraphics[width=0.7in]{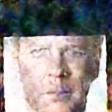}}
\hspace{0mm}
\subfloat[]{\includegraphics[width=0.7in]{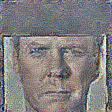}}
\hspace{0mm}
\subfloat[]{\includegraphics[width=0.7in]{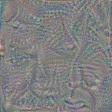}}
\hspace{0mm}
\subfloat[]{\includegraphics[width=0.7in]{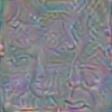}}
\hspace{0mm}
\subfloat[]{\includegraphics[width=0.7in]{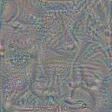}}

\caption{
\justifying
Reconstructed images from facial features and obfuscated facial features generated under white-box and black-box settings. From top to bottom, each of the two rows represents the samples from LFW, AgeDB30, CALFW, and CFP datasets. The images from left to right column are as follows: (a) original images, (b) reconstructed images by RecFace, (c) images optimized through backpropagation under white-box setting, (d) images reconstructed from features with the same amplitude of noises with obfuscated features by RecFace, (e) images reconstructed from features with the same amplitude of noises with obfuscated features under white-box setting, (f) original images after obfuscating based on PMT, (g) images reconstructed by RecFace from the embedding of the obfuscated images, and (h) images optimized based on the embedding of the obfuscated images using backpropagation under white-box setting.}
\label{fig:defense_recons_vis}

\end{figure*}

More strictly, when the attackers are accessible to the shallow model or similar surrogate models, they can conduct an optimization-based reconstruction attack \cite{erdougan2022unsplit}. It is much more challenging to defend. In our experiment, we assume that the attackers are accessible to the knowledge of the user's shallow model, including model structures and parameters. Table \ref{tab:defense_recons} shows that PMT can still work even under the white-box setting. For SRRA, which stands for success rate of reconstruction attack, all reconstructed facial images are highly consistent with the original images on the authorized model. Take the result on CALFW as an example: compared with unprotected samples, SSIM drops from 0.62 to 0.02, PSNR drops from 15.14 to 11.60, while SRRA maintains high accuracy. We also add the same random noises with the same amplitude as the PTM (i.e., the random noises with the same amplitude as the PTM have disrupted the order and added to unprotected features). In comparison, the facial features with random noises have much higher SSIM and PSNR. In particular, from Table \ref{tab:defense_recons}, we can know that the SRRA is always 1.0. The reason is that our idea is to minimize the distance between the processed features by PMT and unprotected features. Compared with the unprotected features, the perturbations of processed features by PMT are small. Meanwhile, the random noise has the same amplitude as the PTM. Hence, the SRRA of the processed features by PMT and adding random noise is always 1.0. From Fig.\ref{fig:defense_recons_vis}, we can know that visualization of the reconstructed facial features with random noises (the amplitude of the noise is set to 0.3), two types of attack can almost reconstruct the facial images with low COS and visually distinguishable facial attributes. These analyses demonstrate that traditional random noises fail to protect our privacy facial data while our PMT stands out.

\subsection{Defense against Data Abuse}
In this subsection, we conduct experiments on defense against data abuse and quantify the performance using Equation \ref{eq:fr_acc}. Note that our goal is to guarantee the performance of authorized models while restricting the performance on other service models. The experiment results involving five models and four datasets are shown in Table \ref{tab:defense_frecog}, where the models in the second column refer to the authorized models, and the models in the first row are applied for evaluation on obfuscated facial data. We can see that our method has little accuracy degradation in comparison with accuracy on the clean data, and the accuracy of unauthorized models is less than 10\% in many conditions, even for models with the same architecture as IR50-Softmax and IR50-Sphere.

\begin{table*}[]
\centering
\caption{Results of defense against unauthorized models on four datasets.}
\label{tab:defense_frecog}
\renewcommand{\arraystretch}{1.2}
\resizebox{1.5\columnwidth}{!}{
\begin{threeparttable}
\begin{tabular}{cccccccc}
\toprule
Datasets & Authorized model & MobileFace & ArcFace & ResNet50 & IR50-Softmax & IR50-Sphereface & Average \\ 
\hline
\multirow{6}{*}{LFW} & Clean & 99.4\% & 98.2\% & 99.2\% & 98.2\% & 99.0\% & 98.8\% \\
 & MobileFace & \underline{99.0\%} & 6.7\% & 8.7\% & 2.9\% & 12.5\% & 7.7\% \\
 & ArcFace & 4.8\% & \underline{100.0\%} & 0.0\% & 2.9\% & 0.0\% & 1.9\% \\
 & ResNet50 & 8.7\% & 4.8\% & \underline{95.2\%} & 16.3\% & 8.7\% & 9.6\% \\
 & IR50-Softmax & 14.4\% & 9.6\% & 1.0\% & \underline{98.1\%} & 12.5\% & 9.4\% \\
 & IR50-Sphereface & 1.9\% & 1.0\% & 1.0\% & 8.7\% & \underline{99.0\%} & 3.2\% \\ \hline
\multirow{6}{*}{AgeDB30} & Clean & 94.9\% & 96.2\% & 96.9\% & 95.7\% & 96.2\% & 96.0\% \\
 & MobileFace & \underline{94.2\%} & 9.6\% & 1.9\% & 5.8\% & 1.9\% & 4.8\% \\
 & ArcFace & 21.2\% & \underline{98.1\%} & 5.8\% & 6.7\% & 3.8\% & 6.7\% \\
 & ResNet50 & 4.8\% & 4.8\% & \underline{93.3\%} & 8.7\% & 10.6\% & 7.2\% \\
 & IR50-Softmax & 17.3\% & 10.6\% & 13.5\% & \underline{97.1\%} & 16.3\% & 14.4\% \\
 & IR50-Sphereface & 3.8\% & 1.9\% & 4.8\% & 3.8\% & \underline{96.2\%} & 3.6\% \\ 
 \hline
\multirow{6}{*}{CALFW} & Clean & 95.9\% & 95.1\% & 95.4\% & 95.9\% & 96.1\% & 95.7\% \\
 & MobileFace & \underline{97.1\%} & 7.7\% & 5.8\% & 6.7\% & 1.0\% & 5.3\% \\
 & ArcFace & 17.3\% & \underline{98.1\%} & 7.7\% & 4.8\% & 1.9\% & 7.9\% \\
 & ResNet50 & 19.2\% & 1.9\% & \underline{96.2\%} & 9.6\% & 8.7\% & 9.9\% \\
 & IR50-Softmax & 21.3\% & 12.5\% & 12.5\% & \underline{99.0\%} & 17.3\% & 15.9\% \\
 & IR50-Sphereface & 4.8\% & 1.9\% & 1.9\% & 6.7\% & \underline{95.2\%} & 3.8\% \\ 
 \hline
\multirow{6}{*}{CFP-FF} & Clean & 99.3\% & 99.3\% & 99.6\% & 99.5\% & 99.4\% & 99.4\% \\
 & MobileFace & \underline{100.0\%} & 1.0\% & 1.0\% & 8.7\% & 8.7\% & 4.9\% \\
 & ArcFace & 7.7\% & \underline{100.0\%} & 2.9\% & 1.0\% & 2.9\% & 3.6\% \\
 & ResNet50 & 13.5\% & 1.0\% & \underline{96.2\%} & 3.8\% & 8.7\% & 6.8\% \\
 & IR50-Softmax & 9.6\% & 3.8\% & 8.7\% & \underline{99.0\%} & 7.7\% & 7.5\% \\
 & IR50-Sphereface & 1.0\% & 1.0\% & 1.9\% & 1.0\% & \underline{98.1\%} & 1.2\% \\
 \bottomrule
\end{tabular}

 \begin{tablenotes}   
    \footnotesize
    \item [1] Note that the second column refers to the authorized models as well as the shallow model on the user side.
    \item [2] \textbf{Clean} means the accuracy of the face recognition model on the original data. 
    \item [3] \textbf{Average} indicates the average accuracy of the unauthorized model.
    \item [4] Underlined fonts indicate the recognition accuracy of the authorized model.
\end{tablenotes}
\end{threeparttable}
}
\end{table*}

\begin{figure}[]
\centering
\subfloat{\includegraphics[width=3in]{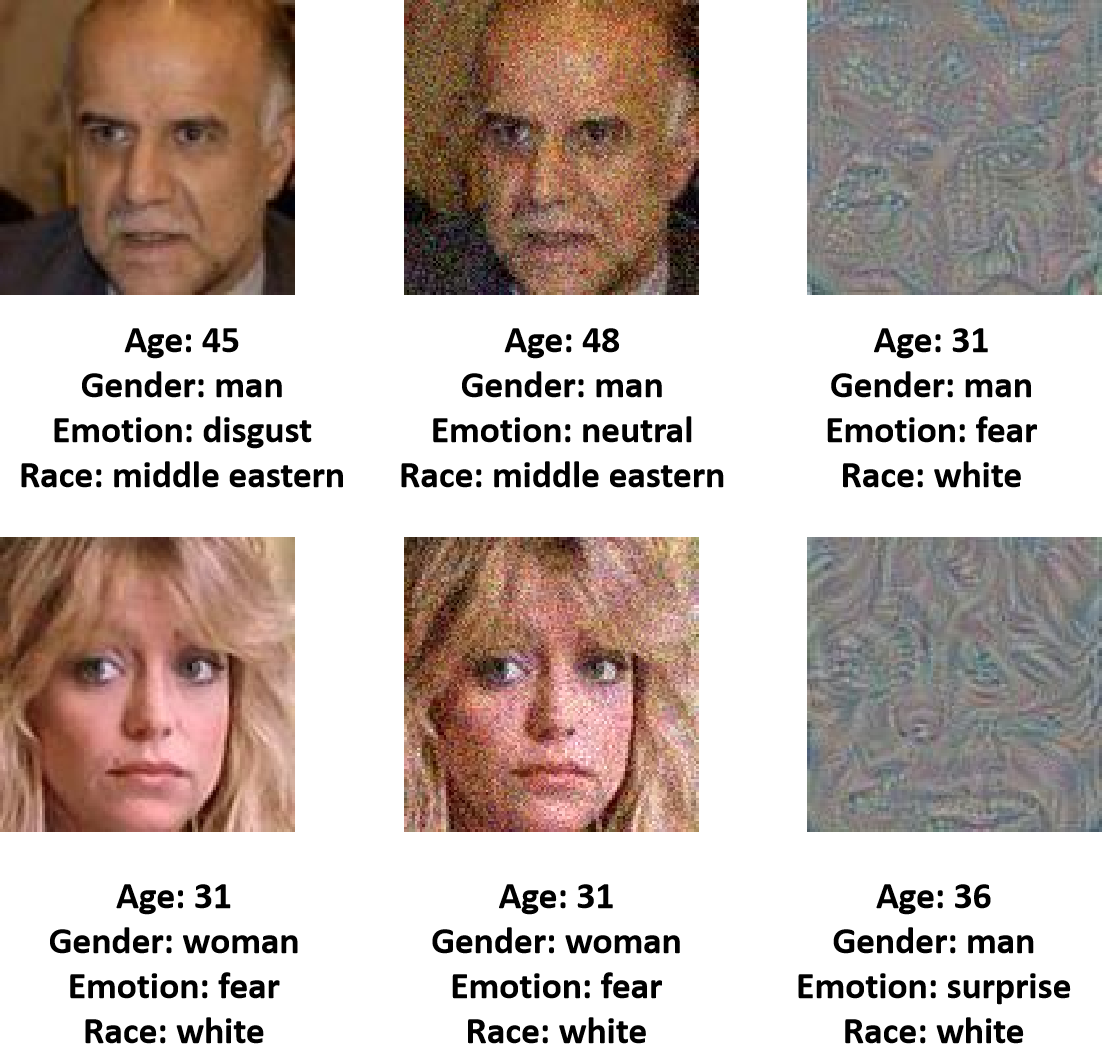}}
\caption{
Examples of attribute estimation attack (The pictures from left to right are: original facial data, facial data with Gaussian noise, obfuscated facial data with PMT). 
}
\label{fig:attributte_res}
\end{figure}

\begin{table}[]
\centering
\caption{Experimental results of attribute estimation attacks.}
\label{tab:defense_attribute}
\renewcommand{\arraystretch}{1.2}
\begin{tabular}{ccllll}
\toprule
Dataset & Images & Emotion$\downarrow$ & Race$\downarrow$ & Gender$\downarrow$ & Age$\uparrow$ \\ 
\hline
\multicolumn{1}{c}{
\multirow{2}{*}{LFW}}
& $\mathrm{PMT}(x)$ & 0.31 & 0.63 & 0.83 & 8.93 \\
\multicolumn{1}{c}{} 
& $x + \delta$ & 0.81 & 0.76 & 0.90 & 5.10 \\ 
\hline
\multicolumn{1}{c}{
\multirow{2}{*}{AgeDB-30}}
& $\mathrm{PMT}(x)$ & 0.32 & 0.76 & 0.63 & 9.63 \\
\multicolumn{1}{c}{}
& $x + \delta$ & 0.88 & 0.87 & 0.85 & 3.72 \\ 
\hline
\multicolumn{1}{c}{
\multirow{2}{*}{CALFW}}
& $\mathrm{PMT}(x)$ & 0.37 & 0.65 & 0.82 & 8.95 \\
\multicolumn{1}{c}{} 
& $x + \delta$ & 0.82 & 0.81 & 0.90 & 5.95 \\ 
\hline
\multicolumn{1}{c}{
\multirow{2}{*}{CFP-FF}}
& $\mathrm{PMT}(x)$ & 0.40 & 0.60 & 0.73 & 8.93 \\
\multicolumn{1}{c}{} 
& $x + \delta$ & 0.84 & 0.87 & 0.94 & 4.20 \\ 
\bottomrule
\end{tabular}

\end{table}

\subsection{Defense against Unauthorized Face Attribute Estimation}
In this setting, we consider that the user needs to upload the original facial images to the server. Therefore, the attackers can steal the facial image and analyze its attributes for advertising recommendation by using an attribute model. To measure the defensive effect of PMT, we use the COS similarity of the attribute vectors between obfuscated and original images. From Table \ref{tab:defense_attribute}, we can know that direct attribute prediction on our obfuscated facial images $\mathrm{PMT}(x)$ suffer from enormous confusion compared with images with Gaussian noises $x + \delta$ (noise amplitude:0.3). More intuitively, Fig.\ref{fig:attributte_res} shows that our PMT can resist to unknown attribute estimation attack effectively to avoid the privacy (age, gender, emotion, race and so on) leakage of the facial image.

\subsection{Trade-off between Privacy-Preserving and Utility}

\begin{figure}[]
\centering
\setcounter{subfigure}{0}
\subfloat[]{\includegraphics[width=1.6in]{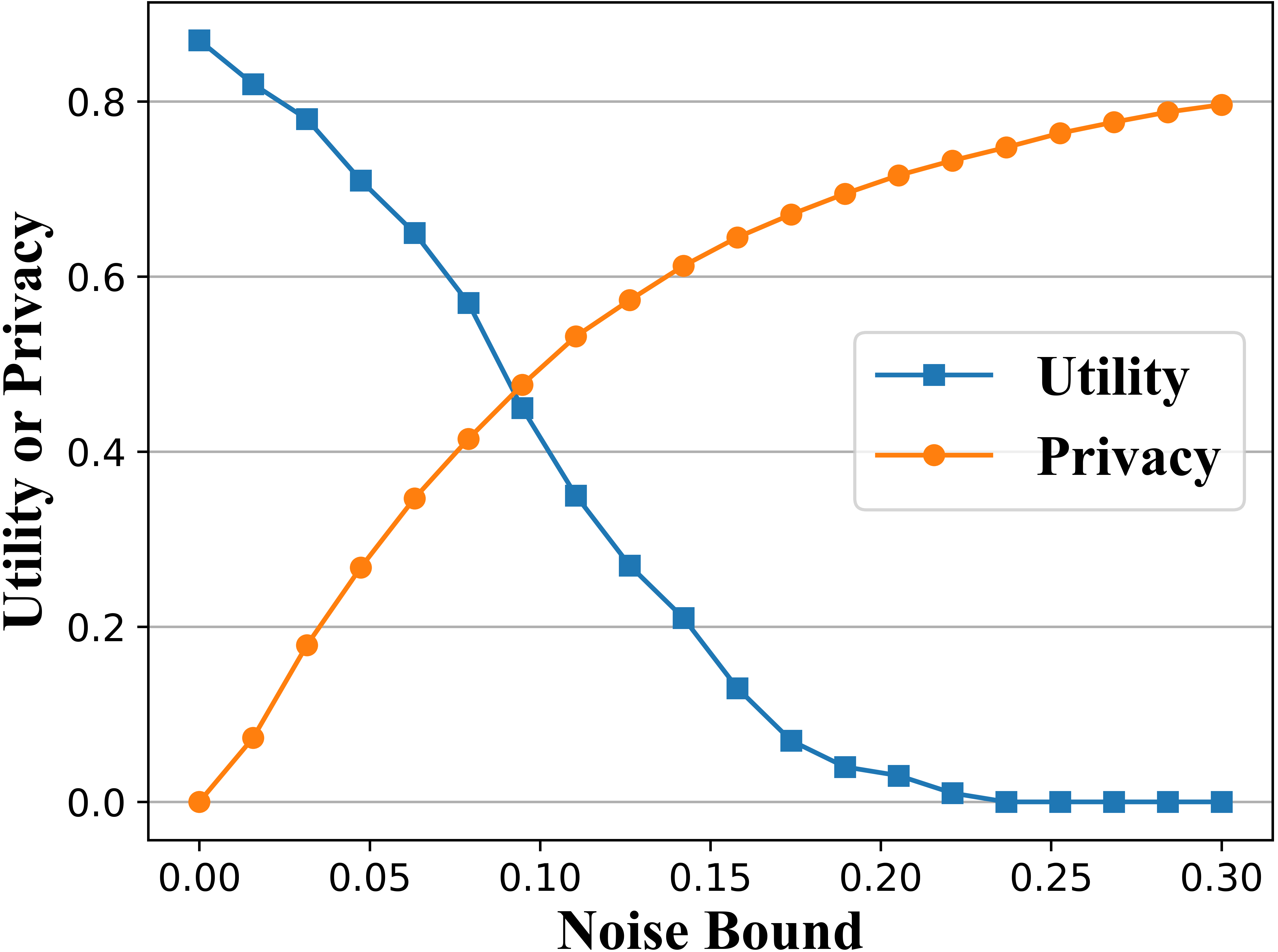}}
\hspace{2mm}
\subfloat[]{\includegraphics[width=1.6in]{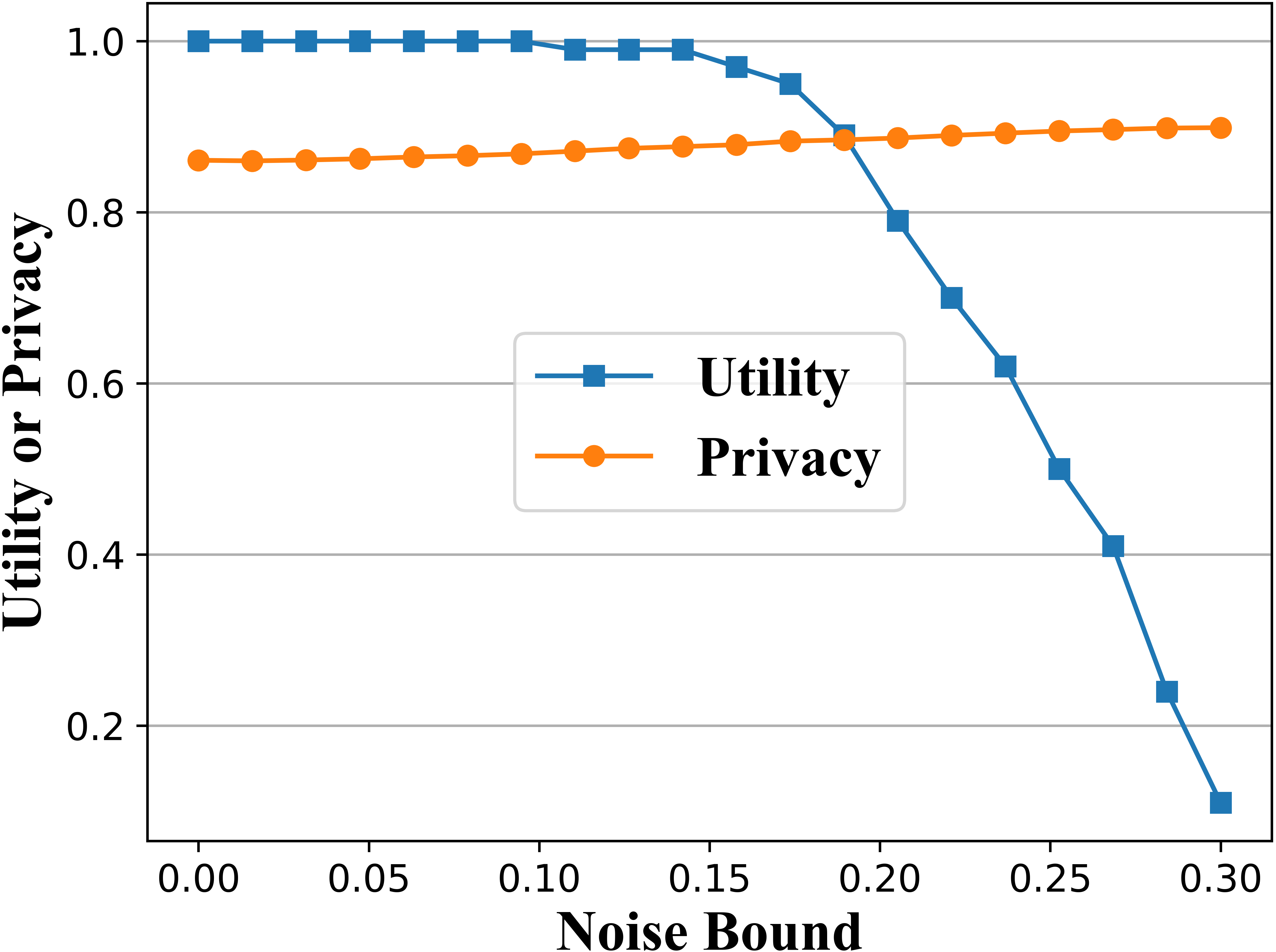}}

\caption{
\justifying
The ``Utility'' and ``Privacy'' of (a) original facial data and (b) obfuscated facial data on LFW.
}
\label{fig:tradeoff}
\end{figure}

We have evaluated the effectiveness of PMT on privacy-preserving (defense against face reconstruction and attribute) and data utility (defense against data abuse) separately. Ideally, we try to find a remarkable balance between privacy-preserving and data utility. We mainly consider the trade-off with variable $\delta$ in $UP(\delta, \kappa)$ ($\kappa$ is the threshold of facial feature similarity and usually set $0.2$ in real-world face recognition service. To show that our method has less impact on the performance of the main task, we set more strict setting: $\kappa = 0.5$). As mentioned in Equation \ref{eq:up}, the first term and second term of $UP(\delta, \kappa)$ denotes the measurement of ``Utility'' and ``Privacy''. To make a more intuitive comparison between the data processed by PMT and the original one, we plotted the change curve of ``Utility'' and ``Privacy'' with noise bound $\delta$ (from 0 to 0.3), as shown in Fig.\ref{fig:tradeoff}. We can see that before the $\delta$ increases to a certain range (from 0 to 0.2), the obfuscated facial data maintain a satisfactory balance and have higher ``Utility'' and ``Privacy'' than the original facial data. 
Only when the $\delta$ is too intensive, there is an obvious degradation of the ``Utility'' with the increase of ``Privacy''. However, the original facial data maintain an unacceptable trade-off between ``Utility'' and ``Privacy''. For example, when the $\delta$ is close to 0.2, the ``Utility'' of the original facial data almost drops to 0.

\begin{figure}[]
\centering
\setcounter{subfigure}{0}
\subfloat[]{\includegraphics[width=1.7in]{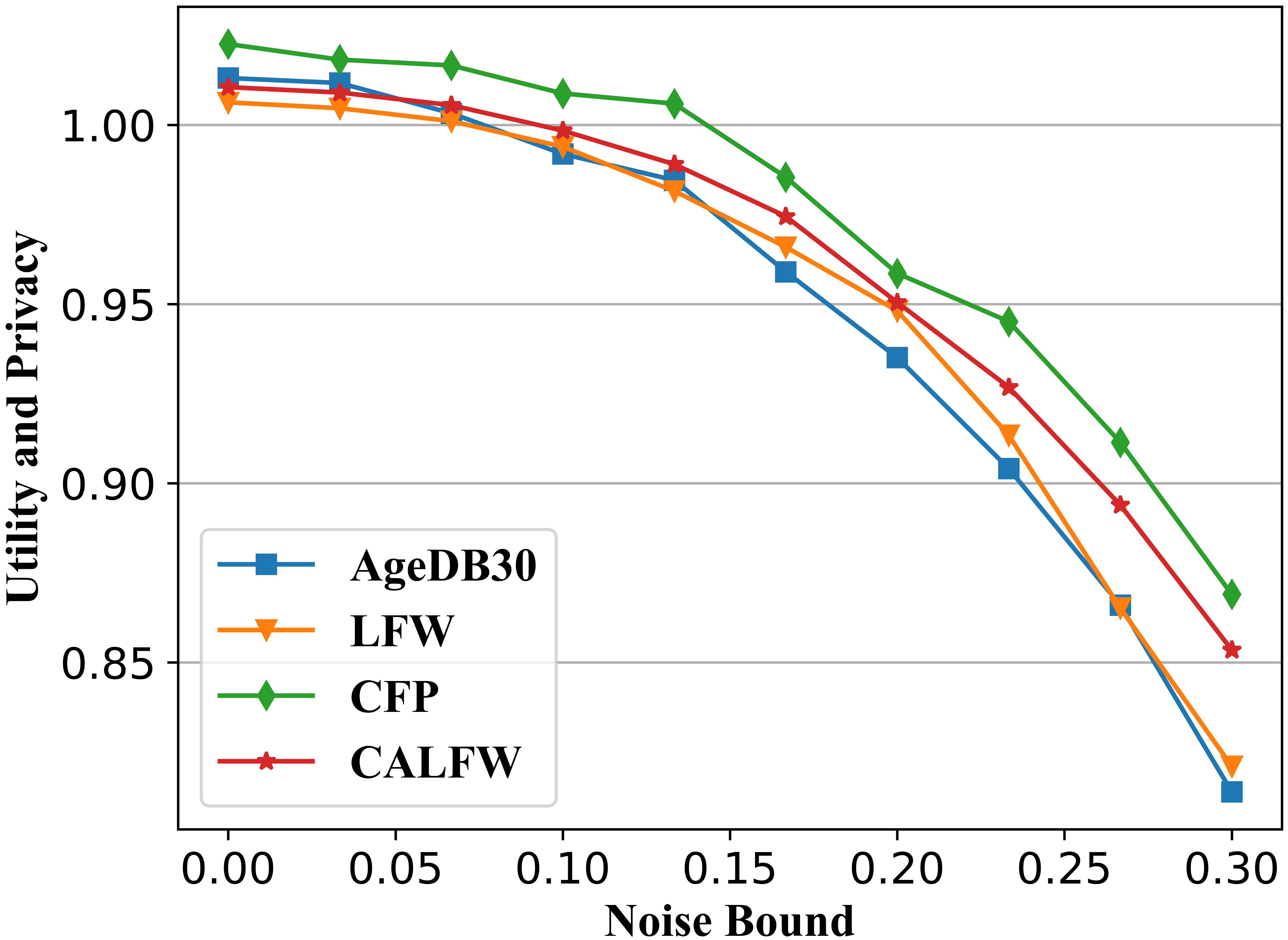}}
\hspace{-1mm}
\subfloat[]{\includegraphics[width=1.7in]{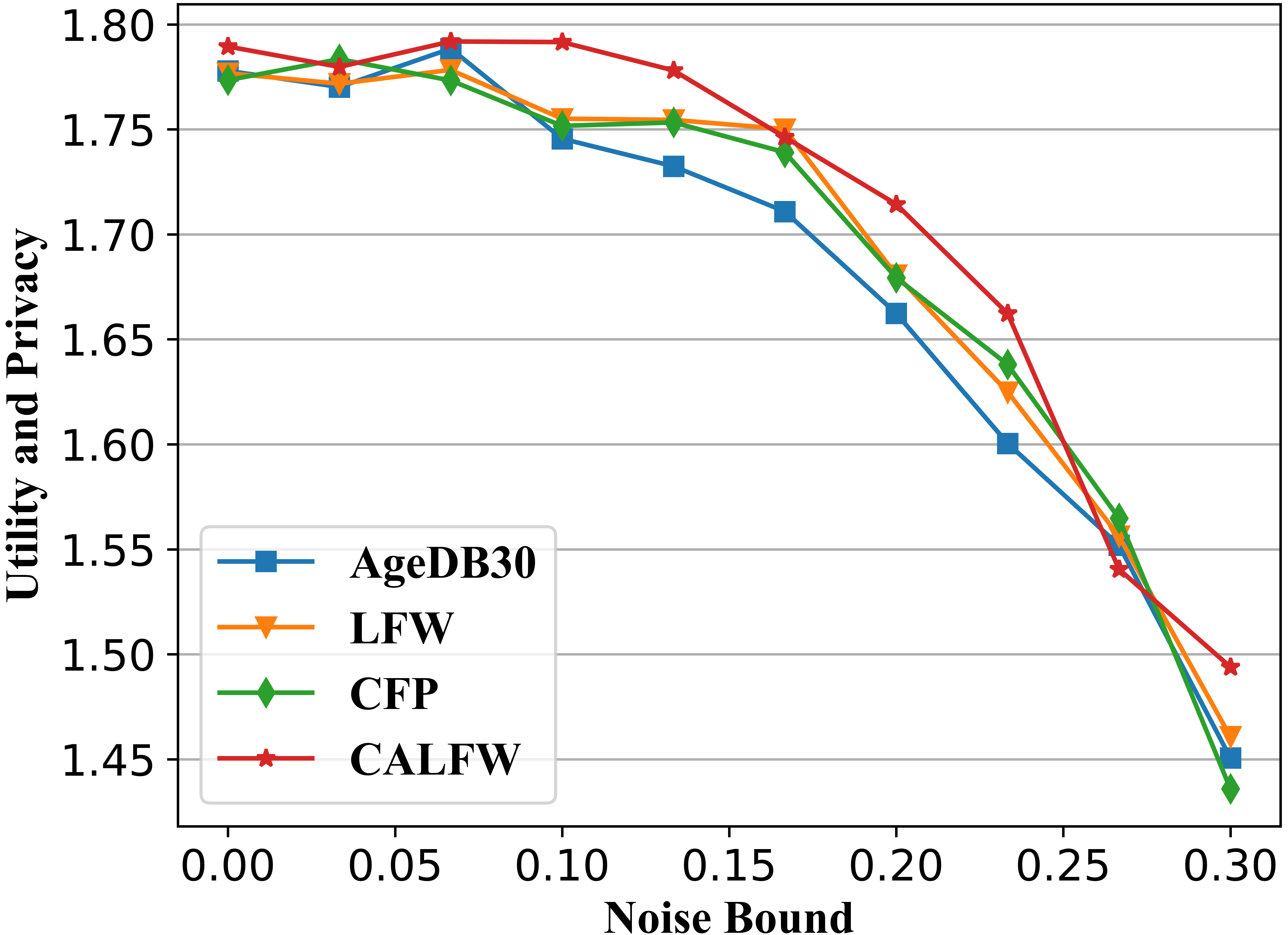}}

\caption{
Performance of PMT with different noise bounds for facial features (left: original features, right: obfuscated features).
}
\label{fig:plot_feature}
\end{figure}

\begin{figure}[]
\centering
\subfloat{\includegraphics[width=3.5in]{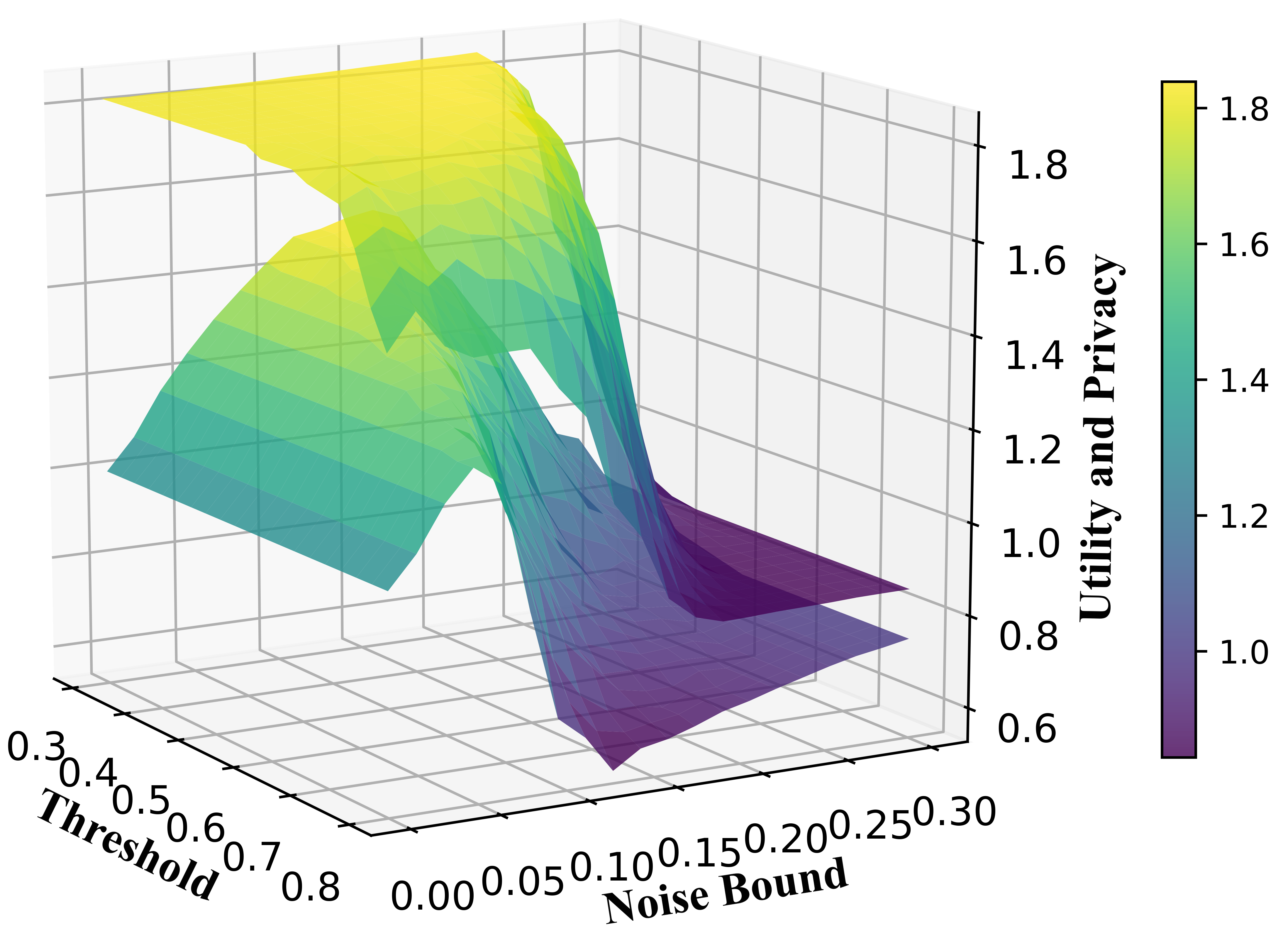}}
\caption{
\justifying
$UP(\delta, \kappa)$ of PMT with different noise bounds $\delta$ and threshold $\kappa$ for facial images on LFW. (The obfuscated facial data is on the top of the figure, while the original facial data is on the bottom.)
}
\label{fig:up_3d}
\end{figure}

\subsection{Robustness of PMT}

\begin{table*}[]
\centering
\caption{Experimental results of scalability for MR-PMT.}
\label{tab:scalability}
\renewcommand{\arraystretch}{1.2}
\begin{threeparttable}
\begin{tabular}{ccccccc}
\toprule
Datasets & Authorized model & \ding{172} & \ding{173} & \ding{174} & \ding{175} & \ding{176} \\
\hline
 \multirow{6}{*}{LFW} 
& \ding{172}+\ding{173} & \underline{99.0\%} & \underline{99.0\%} & 53.8\% & 23.1\% & 20.2\% \\
& \ding{173}+\ding{174} & 63.5\% & \underline{99.0\%} & \underline{99.0\%} & 51.0\% & 28.8\% \\
& \ding{172}+\ding{173}+\ding{174} & \underline{100.0\%} & \underline{99.0\%} & \underline{100.0\%} & 61.5\% & 41.3\% \\
& \ding{173}+\ding{174}+\ding{175} & 64.2\% & \underline{94.2\%} & \underline{100.0\%} & \underline{100.0\%} & 28.8\% \\
& \ding{172}+\ding{173}+\ding{174}+\ding{175} & \underline{99.0\%} & \underline{97.1\%} & \underline{100.0\%} & \underline{99.0\%} & 38.5\% \\
& \ding{173}+\ding{174}+\ding{175}+\ding{176} & 75.2\% & \underline{100.0\%} & \underline{99.0\%} & \underline{100.0\%} & \underline{99.0\%} \\
\midrule
\multirow{6}{*}{AgeDB30}
& \ding{172}+\ding{173} & \underline{93.3\%} & \underline{94.2\%} & 32.7\% & 23.1\% & 22.1\% \\
& \ding{173}+\ding{174} & 54.8\% & \underline{91.3\%} & \underline{96.2\%} & 33.6\% & 21.2\% \\
& \ding{172}+\ding{173}+\ding{174} & \underline{93.3\%} & \underline{97.1\%} & \underline{94.2\%} & 62.5\% & 37.5\% \\
& \ding{173}+\ding{174}+\ding{175} & 51.9\% & \underline{87.5\%} & \underline{91.3\%} & \underline{93.3\%} & 30.7\% \\
& \ding{172}+\ding{173}+\ding{174}+\ding{175} & \underline{88.5\%} & \underline{84.6\%} & \underline{88.5\%} & \underline{89.4\%} & 30.8\% \\
& \ding{173}+\ding{174}+\ding{175}+\ding{176} & 76.5\% & \underline{94.2\%} & \underline{99.0\%} & \underline{95.2\%} & \underline{95.2\%} \\
\midrule
\multirow{6}{*}{CALFW} 
& \ding{172}+\ding{173} & \underline{95.2\%} & \underline{96.2\%} & 51.9\% & 28.8\% & 22.1\% \\
& \ding{173}+\ding{174} & 66.3\% & \underline{95.2\%} & \underline{98.1\%} & 49.0\% & 27.9\% \\
& \ding{172}+\ding{173}+\ding{174} & \underline{95.4\%} & \underline{97.1\%} & \underline{97.1\%} & 73.1\% & 44.2\% \\
& \ding{173}+\ding{174}+\ding{175} & 64.4\% & \underline{94.2\%} & \underline{98.1\%} & \underline{97.1\%} & 29.8\% \\
& \ding{172}+\ding{173}+\ding{174}+\ding{175} & \underline{95.2\%} & \underline{92.3\%} & \underline{96.2\%} & \underline{97.1\%} & 45.2\% \\
& \ding{173}+\ding{174}+\ding{175}+\ding{176} & 74.2\% & \underline{98.1\%} & \underline{98.1\%} & \underline{97.1\%} & \underline{99.0\%} \\
\midrule
\multirow{6}{*}{CFP-FF}
& \ding{172}+\ding{173} & \underline{98.1\%} & \underline{99.0\%} & 24.0\% & 20.2\% & 14.4\% \\
& \ding{173}+\ding{174} & 54.8\% & \underline{97.1\%} & \underline{99.0\%} & 34.6\% & 17.3\% \\
& \ding{172}+\ding{173}+\ding{174} & \underline{100.0\%} & \underline{99.0\%} & \underline{99.0\%} & 55.8\% & 27.0\% \\
& \ding{173}+\ding{174}+\ding{175} & 60.6\% & \underline{96.2\%} & \underline{98.1\%} & \underline{98.1\%} & 15.4\% \\
& \ding{172}+\ding{173}+\ding{174}+\ding{175} & \underline{96.2\%} & \underline{97.1\%} & \underline{98.1\%} & \underline{100.0\%} & 24.0\% \\
& \ding{173}+\ding{174}+\ding{175}+\ding{176} & 75.1\% & \underline{98.1\%} & \underline{98.1\%} & \underline{98.1\%} & \underline{99.0\%} \\
\bottomrule
\end{tabular}
\begin{tablenotes}
\footnotesize
\item [1] Note that \ding{172}\--\ding{176} stand for MobileFace, ResNet50, IR50-Softmax, IR50-Sphereface and ArcFace respectively.
\item [2] The underlined part denotes the accuracy of authorized models.
\end{tablenotes}
\end{threeparttable}
\end{table*}
In real-world applications, the robustness of privacy protection methods becomes exceedingly crucial. The method introduced in this paper deviates from conventional approaches, as it involves perturbing facial features instead of modifying original data. In contrast, the most prevalent data transformation techniques are tailored to original image data. Consequently, when gauging the robustness of PMT, our experiments mainly consider that the preprocessing operation by the service provider is Gaussian noise. Therefore, we evaluate the usability of PMT from two aspects: facial images and facial features.

To evaluate the obfuscated facial feature robustness, we deploy Gaussian noise at different amplitudes. The results of facial features are shown in Fig. \ref{fig:plot_feature}. Compared with the original features, the obfuscated features achieve significantly higher $UP(\delta, \kappa)$. The obfuscated features are also not sensitive to the increase of perturbation amplitude (also known as noise bounds) within a specific range (from 0 to 0.2). 

For evaluating the obfuscated facial image robustness, we further conducted more detailed experiments to discuss the effects of perturbation amplitude and face recognition threshold on $UP(\delta, \kappa)$ respectively. The results are shown in Fig. \ref{fig:up_3d}. Overall, our obfuscated image $\mathrm{PMT}(x)$ outperform the original image. Particularly, before the threshold rises to 0.6, $UP(\delta, \kappa)$ of the obfuscated image remains stable. In contrast, we can notice that the original images always have a peak performance regardless of the amplitude of perturbations, and $UP(\delta, \kappa)$ continues to decline when the threshold increases. The reason is that as the amplitude of perturbations rises, the privacy-preserving performance of original facial data also increases until the noise rises to a certain level and causes a degradation of the utility.

\subsection{Scalability of PMT}
We experimentally validate the scalability of the proposed MR-PMT in this paper. Specifically, we assume that users have authorized face recognition services from multiple providers, involving multi-terminal collaboration during data generation optimization. We assess the performance of the proposed method on both authorized and unauthorized service models based on task's accuracy. The experimental results are shown in Table \ref{tab:scalability}. From Table \ref{tab:scalability}, we can know that the obfuscated data by our method execute on authorized service models with a good performance and a low performance on unauthorized service models. Note that \ding{172}\--\ding{176} stand for MobileFace, ResNet50, IR50-Softmax, IR50-Sphereface and ArcFace respectively. However, we also notice that some unauthorized models have high accuracy (The last row of the experiment for each dataset). We reason that this is because ``MobileFace'' has a small number of parameters. Thus, examples can be easily transferred to these "small models", which is consistent with past studies of adversarial samples\cite{wu2018understanding}.

\subsection{Ablation Study}
\begin{figure}[]
\centering
\subfloat{\includegraphics[width=3in]{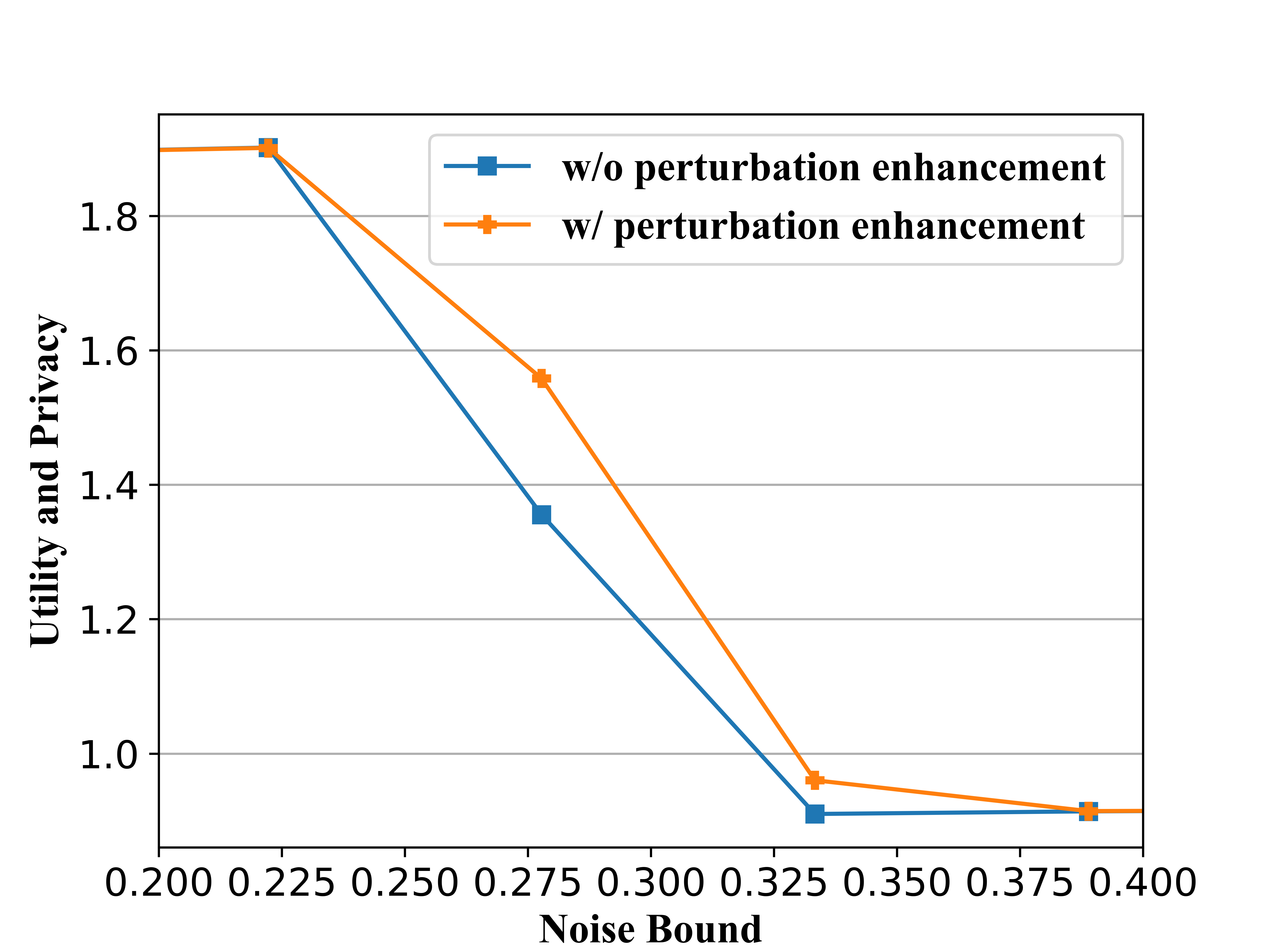}}  
\caption{
Performance of PMT for whether using perturbation enhancement.
}
\label{fig:ablation}
\end{figure}

We analyze the effectiveness of different perturbation enhancement techniques. As shown in Table \ref{tab:ablation_study}, ablation experiments were conducted on LFW (Authorized model: IR50-Softmax, Unauthorized model: MobileFace and Metrics: $UP(0.1,0.3)$). We can find that incorporating all of our techniques has a boost for utility-privacy metrics and the accuracy of the authorized model. For data abuse prevention, our selected combination can achieve the best performance against unauthorized face recognition service. Note that the performance of authorized service has an accuracy decline to an acceptable extent. It may arise from the trade-off of our perturbations between robustness and utility. More intuitively, in Fig.\ref{fig:ablation}, the orange line and blue line represent equipping perturbation enhancement and without it. It is obvious that the proposed perturbation enhancement boosts the performance of PMT with the increase of noise.

\begin{table}[]
\centering
\caption{Experimental results of ablation study on LFW.}
\label{tab:ablation_study}
\renewcommand{\arraystretch}{1.2}
\resizebox{0.8\columnwidth}{!}{
\begin{threeparttable}
\begin{tabular}{cccc}
\toprule
\ding{172}/\ding{173}/\ding{174} & Authorized$\uparrow$ & Unauthorized$\downarrow$ & UP$\uparrow$ \\
\midrule
\ding{56}/\ding{56}/\ding{56} & 98.2\% & 6.9\% & 1.803 \\
\ding{56}/\ding{56}/\ding{52} & 98.1\% & 4.1\% & 1.825 \\
\ding{56}/\ding{52}/\ding{56} & 98.1\% & 5.7\% & 1.822 \\
\ding{52}/\ding{56}/\ding{56} & 98.7\% & 6.3\% & 1.812 \\
\ding{52}/\ding{52}/\ding{56} & 99.0\% & 5.8\% & 1.831 \\
\ding{52}/\ding{56}/\ding{52} & 99.4\% & 2.8\% & 1.829 \\
\ding{56}/\ding{52}/\ding{52} & 99.0\% & 3.9\% & 1.831 \\
\ding{52}/\ding{52}/\ding{52} & 99.0\% & 1.0\% & 1.848 \\
\bottomrule
\end{tabular}
\begin{tablenotes}
\footnotesize
\item Note that \ding{172}\--\ding{174} stand for our perturbation enhancement techniques: data augmentation, translation-invariant enhancement, and layer aggregation. 
\end{tablenotes}
\end{threeparttable}
}
\end{table}

\section{Conclusion}

In this work, we proposed a PMT method to address data abuse and privacy leakage concerns in face recognition applications. This method performs privacy minimization transformation before data is transmitted to the service provider, preventing user's data from illegal privacy theft and abuse by unauthorized entities. Extensive experimental results show the superior performance of PMT in defending against data abuse, reconstruction attacks, and attribute estimation attacks. Moreover, the experiments also validate that PMT can exhibit remarkable robustness and scalability and achieve a desirable trade-off between privacy-preserving and utility. It positions our method with significant potential and crucial practicality for real-world application scenarios.

\bibliographystyle{IEEEtran}  
\bibliography{references}
\end{document}